\documentclass[aps,preprint,nofootinbib,superscriptaddress]{revtex4}
\usepackage{color}
\usepackage{longtable}
\usepackage{graphicx}
\usepackage{amsmath}
\usepackage{amsfonts}
\usepackage{amssymb}
\usepackage{natbib} % Use the natbib bibliography and citation package
\usepackage{setspace}
\usepackage{xspace}
\usepackage{cancel}
\usepackage{physics}
\usepackage{tensor,braket,color,bbold,dsfont,slashed}
\usepackage{mathrsfs}

\usepackage[normalem]{ulem}
\renewcommand\sout{\bgroup\color[rgb]{0,0.4,0.8} \ULdepth=-.5ex \ULset}

\begin{document}
\title{Effect of nucleon effective mass and symmetry energy on the neutrino mean free path in a neutron star}

\author{Parada T. P. Hutauruk}%\email{phutauruk@gmail.com}
\affiliation{Department of Physics, Pukyong National University, Busan 48513, Korea}
\affiliation{Department of Physics Education, Daegu University, Gyeongsan 38453, Korea}

\author{Hana Gil}%\email{khn1219@gmail.com}
\affiliation{Center for Extreme Nuclear Matters, Korea University, Seoul 02841, Korea}

\author{Seung-il Nam}\email{sinam@pknu.ac.kr}
\affiliation{Department of Physics, Pukyong National University, Busan 48513, Korea}
\affiliation{Center for Extreme Nuclear Matters, Korea University, Seoul 02841, Korea}
         
\author{Chang Ho Hyun}%\email{hch@daegu.ac.kr}
\affiliation{Department of Physics Education, Daegu University, Gyeongsan 38453, Korea}
\affiliation{Center for Extreme Nuclear Matters, Korea University, Seoul 02841, Korea}

\date{\today}

\begin{abstract}
The Korea-IBS-Daegu-SKKU energy density functional (KIDS-EDF) models, constructed from the perturbative expansion of the energy density in nuclear matter, have been successfully and widely applied in describing the properties of finite nuclei and infinite nuclear matter. In the present work, we extend the applications of the KIDS-EDF models to investigate the implications of the nucleon effective mass and nuclear symmetry energy for the properties of neutron stars (NSs) and neutrino interaction with the NS constituent matter in the linear response approximation (LRA). At fixed neutrino energy and momentum transfer, we analyze the total differential cross section of neutrino, the neutrino mean free path (NMFP), and the NS mass-radius (M-R) relations. Remarkable results are given by the KIDS0-m*87 and SLy4 models, in which $M_n^* /M \lesssim 1$, and their NMFPs are quite higher in comparison with those obtained from the  KIDS0, KIDS-A, and KIDS-B models, which result in $M_n^*/M \gtrsim 1$. For the KIDS0, KIDS-A, and KIDS-B models, we obtain $\lambda \lesssim R_{\textrm{NS}}$, indicating that these models could predict the slow NS cooling and neutrino trapping in NSs. In contrast, the KIDS0-m*87 and SLy4 models yield $\lambda \gtrsim R_{\textrm{NS}}$ and thus we expect faster NS cooling and a small possibility of neutrino trapping within NSs.  We also calculate the NMFP as a function of the neutrino energy and the nuclear matter density and find that the NMFP decreases as the density and neutrino energy increase, which is consistent with the result obtained in the Brussels-Montreal Skyrme (BSk17 and BSk18) models at saturation density. 
\end{abstract}
\maketitle
%------------------------------------------
\section{Introduction} \label{sec:intro}
%------------------------------------------
%
It is widely known that neutron stars (NSs) are compact objects which are extraordinary laboratories for dense stellar matter physics that cannot be reproduced in terrestrial laboratories. The neutrino processes in the later stage of stellar evolution play an important role in supernova explosions, formation of proto-neutron stars (PNSs), and the cooling of NSs~\cite{Baym:1978jf,Lattimer:2004pg,Bethe:1990mw}. However, the internal composition of the cores of NSs and the interactions among the NS constituent matter are still poorly understood. Also, the most important observable properties of NSs, such as NS maximum masses and radii, are not well-constrained yet~\cite{Galloway:2007dn,Lattimer:2006xb}. The NS maximum mass is an implication of general relativity and it is controlled by the equation of state (EoS) of nuclear matter at densities more than a few times normal density. In addition to the determinations of the NS mass and radius, it is expected that the interior of NS will cool via the neutrino emission process, which sensitively depends on the NS composition and the interactions between the NS constituents. Slow or rapid neutrino emissions have a significant implication for the NS cooling process. The neutrino scattering in dense nuclear matter and their emission from the NS are sensitive to the nuclear symmetry energy of the EoS and the nucleon effective masses \cite{Suleimanov:2010th,Yakovlev:2000jp}.

The nucleon effective masses, which are calculated from the effective interaction in the nuclear medium~\cite{Jaminon:1989wj}, are not only crucial for neutrino scattering and absorption (opacity) in  NS~\cite{Hutauruk:2021cgi,Reddy:1997yr,Niembro:2001hd,Hutauruk:2020mhl,Hutauruk:2018cgu} but also for the structure of rare isotopes, 
stellar matter, compact stars, and other astrophysical objects, i.e., supernovae and NSs as well as the dynamics of heavy-ion collisions (HICs)~\cite{Russotto:2016ucm}. Besides the nucleon effective masses, the nuclear symmetry energy also has an essential role in determining the features of the stiffness or the softness of the EoS in neutron-rich matter. 
Generally, stiffness of the nuclear symmetry energy can be understood from the enhancement or reduction of the pressure gradient in the asymmetric matter. Therefore, the nuclear symmetry energy is a powerful tool for controlling the rate of the NS cooling process in a mixture of nucleons, determining the density for the appearance of hyperons or other exotic particles, the nucleon emissions in the reaction dynamics, and the collective flows in HICs~\cite{Li:2002qx,Li:1996ix}.

The behavior of the density dependence of the nuclear symmetry energy at high densities remains unknown~\cite{Li:2019xxz,Baldo:2016jhp}. Although the nuclear symmetry energy at saturation density $\rho_0 \simeq $ 0.16 fm$^{-3}$ was predicted to be around 32 $\pm$ 0.59 MeV in a recent analysis using the nuclear liquid drop (LD) model~\cite{Cao:2022kny}, its values at baryon densities larger than the saturation densities are poorly known~\cite{Li:2019xxz,Baldo:2016jhp}. 
Therefore, intensive studies on the density dependence of the nuclear symmetry energy at higher densities are needed to gain a better understanding of the properties of a NS and its matter constituent interactions as well as to constrain the EoS of neutron-rich matter. Several theoretical works using various microscopic theories with realistic nucleon-nucleon forces and phenomenological models~\cite{Zuo:2005hw,vanDalen:2005ns,Li:2019xxz,Baldo:2016jhp,Cao:2022kny} have been done to observe the features of the
nuclear symmetry energy and nucleon effective masses. 

Besides those theoretical attempts and terrestrial laboratories constraints which allow us to constrain the nuclear symmetry energy at around saturation density, nowadays it is also possible to constrain the nuclear symmetry energy at higher density via the well-known mass-radius (M-R) relations of NSs using astrophysical or astronomical observations such as those from the Neutron Star Interior Composition Explorer (NICER)~\cite{Miller:2019cac,Riley:2019yda}, the Laser Interferometer Gravitational-Wave Observatory (LIGO)/Virgo~\cite{LIGOScientific:2017vwq,LIGOScientific:2018cki}, and other x-ray burst observations~\cite{Tanvir:2013pia}. For instance, 
the astrophysical observation information of the tidal deformability detected from binary NS mergers allows us to constrain the density dependence of nuclear symmetry energy. 
Motivated by recent progress in theoretical studies, terrestrial laboratories or experiments, and astrophysical observations, in the present theoretical study we calculate the neutrino mean free paths (NMFPs) in various Korea-IBS-Daegu-SKKU energy density functional (KIDS-EDF) models and study their relation with the NS masses and radii obtained from observation constraints, which are crucial quantities for the explanation of NS cooling and an important input for the simulation of the neutrino transport~\cite{Rizzo:2005mk}.

In this paper, we investigate the effect of the nucleon effective masses and nuclear symmetry energies on the properties of NSs and on neutrino scattering with NS constituent matter, 
which consists of protons ($p$), neutrons ($n$), electrons ($e$) and muons ($\mu$) as the standard matter particles.  In the present work, selected models are KIDS0, KIDS0-m*87, KIDS-A, KIDS-B, and SLy4. Rules to select the models will be discussed in detail in Sec.~II. In the neutrino scattering, we employ the linear response approximation (LRA) to describe the interaction between neutrino and NS constituents. Note that, in this work, we focus on only the neutrino interactions with the neutrons and protons, since the differential cross sections (DCRS) of the neutrino and the NMFP are more dominant than those of leptons. The KIDS-EDF models have been successfully and widely applied in many physics phenomena such as finite nuclei~\cite{Choi:2021zbc}, quasi elastic electron scattering~\cite{Gil:2021dgd}, inclusive electron scattering~\cite{Gil:2021twe}, and nuclear matter~\cite{Choi:2021zbc} as well as NSs~\cite{Gil:2021ols}. The nuclear symmetry energy, the neutron and proton effective masses, and the neutron and proton fractions of constituents of $\beta$-stable matter are calculated within the KIDS-EDF models. With these quantities obtained from the KIDS-EDF models, we calculate the NMFP in a NS, which is an inverse of the DCRS of the neutrino. We also calculate the M-R relations of NSs by solving the Tolman-Oppenheimer-Volkoff (TOV) equations~\cite{Tolman:1939jz,Oppenheimer:1939ne} and observe the relations between the NS M-R and the NMFP, which are expected to provide new insight into the rate of the NS cooling process and the possibility of neutrino trapping in NSs.

This paper is organized as follows. In Sec.~\ref{sec:kids-edf} we briefly introduce the formalism of the KIDS-EDF models and summarize the parameters used in the models. In Sec.~\ref{sec:MRrelation}, we calculate the neutron star matter EoS by solving the baryon number conservation, charge neutrality, and $\beta$-equilibrium conditions consistently. As a result,  the neutron and proton effective masses and their particle fractions within the NS are obtained. Solving the TOV equations, we obtain the M-R relations of the selected models. In Sec.~\ref{sec:NMFP} we present the DCRS of the neutrino and the NMFP using the linear response approach, and their relations with the NS M-R properties. Section~\ref{sec:summary} is devoted to a summary and conclusion.

%------------------------------------------
\section{KIDS-EDF Model} \label{sec:kids-edf}
%------------------------------------------
%

In this section, we present the formalism of the KIDS-EDF models, which were proposed for the first time for homogeneous nuclear matter~\cite{Papakonstantinou:2016zpe}. They were then applied for describing the properties of finite nuclei. In finite nuclei, the EDF in the nuclear matter was transformed to the form of the Skyrme functional, where the specific values of the effective masses could be reproduced by adjusting the parameters of the functional. Here, we first describe the energy per nucleon in the homogeneous nuclear matter by expanding the energy per nucleon or the energy density in terms of the power of the Fermi momentum $k_F$, which is equivalent to the cubic root of the baryon density at zero temperature. An explicit expression of the KIDS-EDF energy per nucleon is given by
%-----------------
\begin{eqnarray}
  \label{eq1}
  \mathcal{E} \left(\rho, \delta \right) = \mathscr{T} \left(\rho, \delta \right) + \sum_{j=0}^{3} c_{j} (\delta) \rho^{(1+ a_j)},
\end{eqnarray}
where $a_j = j/3$ and $\rho = \rho_n + \rho_p$ is the baryon density. $\rho_n$ and $\rho_p$ are respectively neutron and proton densities. The isospin asymmetry parameter is defined by $\delta = (\rho_n - \rho_p)/\rho$, where $\delta = 0$ for symmetric nuclear matter (SNM) and $\delta = 1$ for pure neutron matter (PNM). The kinetic energy in the first term $\mathscr{T}  \left(\rho, \delta \right)$ of Eq.~(\ref{eq1}) is given by
%---------------
\begin{eqnarray}
  \label{eq2}
  \mathscr{T} \left( \rho, \delta \right) &=& \frac{3}{5} \left[ \frac{\hbar^2}{2M_p} \left( \frac{1-\delta}{2} \right)^{\frac{5}{3}} +  \frac{\hbar^2}{2M_n} \left( \frac{1+\delta}{2} \right)^{\frac{5}{3}}\right] \left( 3\pi^2 \rho \right)^{\frac{2}{3}},
\end{eqnarray}
where $M_p$ and $M_n$ are respectively the proton and neutron masses in free space. The potential energy term in Eq.~(\ref{eq1}) contains the parameters $c_j (\delta)$ which are defined by $c_j (\delta) = \alpha_j + \delta^2 \beta_j$, where the parameters $\alpha_j$ are to be fixed by the SNM properties and $\beta_j$ could be fixed by the EoS of asymmetric nuclear matter. Further details about determining the coefficients of $\beta_j$ and $\alpha_j$ can be found in Ref.~\cite{Gil:2019qfr}.

With the energy per nucleon in Eq.~(\ref{eq1}) expanded in terms of $\delta$, the nuclear symmetry energy $\mathcal{S}(\rho)$ is straightforwardly determined $\textit{via}$ the second derivative of the energy density over the $\delta$. The expression of the symmetry energy $\mathcal{S}(\rho)$ is given by
\begin{eqnarray}
  \label{eq2a1}
  \mathcal{E}(\rho,\delta) &=& \mathcal{E}(\rho, \delta = 0) + \mathcal{S} (\rho) \delta^2 + \mathcal{O} (\delta^4), \nonumber\\
  \mathcal{S}(\rho) &=&\frac{\hbar^2}{6 M} \left( \frac{3\pi^2}{2} \right)^{\frac{2}{3}} \rho^{\frac{2}{3}} + 
\sum_{j=0}^{3} \beta_j\rho^{(1+ a_j)},
\end{eqnarray}
where $ \frac{\hbar^2}{6 M} \left( \frac{3\pi^2}{2} \right)^{\frac{2}{3}} \rho^{\frac{2}{3}}$ is the kinetic energy term, and we replace $M_n$ and $M_p$ with the average nucleon mass $M=(M_n+M_p)/2$. At around the nuclear matter saturation density, the energy per nucleon in SNM $\mathcal{E}(\rho, 0)$ and the nuclear symmetry energy can be respectively expanded by
\begin{eqnarray}
  \label{eq2a2}
  \mathcal{E} (\rho, 0) &=& E_0 + \frac{1}{2} K_0 \mathcal{X}^2 + \frac{1}{6} Q_0 \mathcal{X}^3  + \mathcal{O} (\mathcal{X}^4), \nonumber \\
  \mathcal{S} (\rho) &=& J + L \mathcal{X} + \frac{1}{2}K_\mathrm{sym} \mathcal{X}^2 + \frac{1}{6} Q_\mathrm{sym} \mathcal{X}^3 + \frac{1}{24} R_\mathrm{sym} \mathcal{X}^4 + \mathcal{O} (\mathcal{X}^5),
\end{eqnarray}
where $\mathcal{X} = \left( \frac{\rho-\rho_0}{3\rho_0} \right)$. The compression modulus $K_0$ and the skewness coefficient $Q_0$ in Eq.~(\ref{eq2a2}) are respectively given by
\begin{eqnarray}
  \label{eq2a3}
  K_0 = 9 \rho_0^2 \frac{d^2\left[ \mathcal{E} (\rho,0)/\rho \right]}{d\rho^2} \Big|_{\rho = \rho_0}, \,\,\,\,
  Q_0 = 27 \rho_0^3  \frac{d^3 \mathcal{E} (\rho,0)}{d\rho^3} \Big|_{\rho = \rho_0}.
\end{eqnarray}

Density dependence of the symmetry energy around the saturation density is specified by the value of $J = \mathcal{S} (\rho_0)$, the slope $L$, the curvature $K_\mathrm{sym}$, the skewness $Q_\mathrm{sym}$, and the kurtosis $R_\mathrm{sym}$ which are given  respectively by
\begin{eqnarray}
  \label{eq2a4}
  L = \rho_0 \frac{d \mathcal{S} (\rho)}{d\rho} \Big|_{\rho = \rho_0}, \,\,\,\,
  K_\mathrm{sym} = 9\rho_0^2 \frac{d^2 \mathcal{S} (\rho) }{d\rho^2} \Big|_{\rho = \rho_0},
  \cr
  Q_\mathrm{sym} =27 \rho_0^3 \frac{d^3 \mathcal{S} (\rho)}{d\rho^3} \Big|_{\rho = \rho_0}, \,\,\,\,
  R_\mathrm{sym} = 81 \rho_0^4 \frac{d^4 \mathcal{S} (\rho)}{d\rho^4} \Big|_{\rho = \rho_0}.
\end{eqnarray}

We now turn to determine the KIDS-EDF parameters in terms of the Skyrme force parameters. The conventional Skyrme interaction is defined by~\cite{sly4}
\begin{eqnarray}
  \label{eq2a}
  \mathcal{V}_{i,j} (\mathbf{k},\mathbf{k}') &=& t_0 (1 + x_0 P_\sigma ) \delta ( \mathbf{r}_i - \mathbf{r}_j) + \frac{1}{2} t_1 (1 + x_1 P_\sigma ) \left[ \delta (\mathbf{r}_i - \mathbf{r}_j)\mathbf{k}^2 + \mathbf{k}^{'2} \delta (\mathbf{r}_i - \mathbf{r}_j )\right] \nonumber \\
  &+& t_2 (1 + x_2 P_\sigma ) \mathbf{k}' \cdot \delta (\mathbf{r}_i - \mathbf{r}_j ) \mathbf{k} + \frac{1}{6} t_{3} (1 + x_3 P_\sigma) \rho^{\alpha} \delta (\mathbf{r}_i - \mathbf{r}_j) \nonumber \\
  &+& i W_0 \mathbf{k}' \times \delta (\mathbf{r}_i - \mathbf{r}_j)\mathbf{k} \cdot (\mathbf{\sigma}_i - \mathbf{\sigma}_j),
\end{eqnarray}
where $P_\sigma = \left(1 + \sigma_1 \cdot \sigma_2 \right)/2$ and $\sigma$ are respectively the spin-exchange operator and Pauli spin matrices. 
$\mathbf{k} = (\nabla_i - \nabla_j) /2i$ and $\mathbf{k}' = (\nabla_i^{'} - \nabla_j^{'}) /2i$ are the relative momenta operating in the initial and final states, respectively.
The strength of the spin-orbit coupling $W_0$ is absent in the EDF for homogeneous matter in the Skyrme force. The energy density functional for infinite nuclear matter can be written in terms of the Skyrme force parameters as
\begin{eqnarray}
  \label{eq2b}
  \mathcal{E} (\rho, \delta) &=& \mathscr{T} (\rho, \delta) + \frac{3}{8} t_0 \rho - \frac{1}{8} (2y_0 + t_0) \rho \delta^2 + \frac{1}{16} t_{3} \rho^{(\alpha +1)} \nonumber \\
  &-& \frac{1}{48} (2y_{3} + t_{3} ) \rho^{(\alpha +1)} \delta^2 + \frac{1}{16} (3t_1 + 5 t_2 + 4 y_2) \tau \nonumber \\
  &-& \frac{1}{16} \left[ (2 y_1 + t_1) - (2y_2 + t_2 )\right] \tau \delta^2,
\end{eqnarray}
where $y_i \equiv t_i x_i$. Matching Eqs.~(\ref{eq1}) and~(\ref{eq2b}), we then determine the relations between $c_j(\delta)$ and Skyrme coefficients ($t_i,y_i)$ and one has
\begin{eqnarray}
  \label{eq2c}
  c_0 (\delta) &=& \frac{3}{8} t_0 - \frac{1}{8} (2y_0 + t_0) \delta^2, \nonumber \\
  c_1 (\delta) &=& \frac{1}{16} t_{31} - \frac{1}{48} (2 y_{31} + t_{31} ) \delta^2, \nonumber \\
  c_2 (\delta) &=& \frac{1}{16} t_{32} - \frac{1}{48} (2 y_{32}+ t_{32}) \delta^2 \nonumber \\
&+& \frac{3}{5} \left( \frac{6\pi^2}{\nu} \right)^{\frac{2}{3}} \frac{1}{16} \left\{ (3t_1 + 5t_2 + 4y_2) -\left[(2y_1 + t_1) -(2y_2 + t_2)\right] \delta^2 \right\}, \nonumber \\
  c_3 (\delta) &=& \frac{1}{16} t_{33}- \frac{1}{48} (2y_{33}+ t_{33}) \delta^2,
\end{eqnarray}
where the power of density $\alpha = 1/3$ is assigned to $t_{31}$ and $y_{31}$, $\alpha = 2/3$ to $t_{32}$ and $y_{32}$, and $\alpha =1$ to $t_{33}$ and $y_{33}$. 
$\nu$ is the spin and isospin degeneracy factor with $\nu=4$ for SNM and $\nu=2$ for the asymmetric matter. As mentioned before, we expand the EDF as a power series of the cubic root of nuclear density. So, in this case, we have three extra terms rather than a single density dependence $\rho^{\alpha}$ in the conventional Skyrme force. Note that, in the KIDS-EDF model, we have $13$ parameters in total to be determined in the Skyrme force. As mentioned earlier, in the present work, we consider the various KIDS-EDF models: KIDS0, KIDS0-m*87, KIDS-A, KIDS-B, and SLy4.

In the KIDS0 model, $\alpha_0$, $\alpha_1$ and $\alpha_2$ are adjusted to three SNM data: $\rho_0=0.16$~fm$^{-3}$, $E_{\rm B} = 16$~MeV, and $K_0 = 240$~MeV;
and four $\beta_i$'s are fitted to a PNM EoS calculated with an {\it ab initio} nuclear force \cite{apr}. There are no presumed values for determining the effective mass, so for the KIDS0 model we obtain isoscalar effective mass $m^*_s \simeq 1.0 M$ and isovector one $m^*_v \simeq 0.8 M$ as a result of the parameter fitting. Considering the effective masses of $m^*_s \simeq 0.8M$ and $m^*_v \simeq 0.7 M$, which are set based on the empirical constraint values~\cite{ppnp2018}, the Skyrme parameters are calibrated to reproduce these effective mass values in the KIDS0-m*87 model. Since both KIDS0 and KIDS0-m*87 models are adjusted to identical nuclear matter properties, the two models have the same symmetry energy parameters. The symmetry energy of the KIDS-A and KIDS-B models are also determined to satisfy the recent neutron star observation constraint~\cite{npsm2021}. These models are within the uncertainty of the symmetry energy parameters consistent with the neutron star data~\cite{Gil:2021ols}. Therefore, the KIDS-A model has different symmetry energy parameters from those in the KIDS-B model. In addition, the specific values of the effective masses are not assumed in the KIDS-A and KIDS-B models, and their obtained isoscalar and isovector effective masses are quite similar to those of the KIDS0 model. Therefore, the difference between the KIDS0, KIDS-A, and KIDS-B models clearly shows the effect of the symmetry energy. Oppositely, the KIDS0, KIDS0-m*87, and SLy4 models have rather similar symmetry energy parameters, so their difference will be obviously clarified by the effective mass. 

Using the standard Skyrme force~\cite{sly4}, the nucleon effective masses in the asymmetric nuclear matter (ANM) are defined by
\begin{eqnarray}
M^*_i = M_i \left[ 1 + \frac{M_i}{8 \hbar^2} \rho \Theta_s - \frac{M_i}{8 \hbar^2} \tau^i_3 \left( 2 \Theta_v - \Theta_s \right) \rho \delta \right]^{-1},
\label{eq:efmn}
\end{eqnarray}
where $M_i$ is the nucleon mass in free space ($i=n,\, p$). The Skyrme force parameters are given as $\Theta_s = 3t_1 + (5t_2+4y_2)$ and $\Theta_v=(2 t_1 +y_1) + (2 t_2 +y_2)$, where the values of parameters of $t_1$, $t_2$, $y_1$, and $y_2$ are given in Table~\ref{tab1}. $\tau_3$ is the third component of the isospin of a nucleon, with $\tau_3 = +1, \, -1$ for the neutron and proton, respectively.  With the Skyrme force parameters of $\Theta_s$ and $\Theta_v$, the nucleon isoscalar and isovector mass ratios 
are defined by
\begin{eqnarray}
  \label{eq3a1}
  \mu_s^* =m^*_s/M= \left( 1 + \frac{M}{8 \hbar^2} \rho \Theta_s \right)^{-1}, \,\,\,\,
  \mu_v^* =m^*_v/M = \left( 1 + \frac{M}{4 \hbar^2} \rho \Theta_v \right)^{-1}.
\end{eqnarray}
Here we notice that the nucleon effective masses can be also defined in terms of the isoscalar and isovector masses {through Eqs.~(\ref{eq:efmn}) and~(\ref{eq3a1}). 
For completeness, the Skyrme force parameters, isoscalar and isovector effective masses, $K_0$, $J$, and $L$ values for the KIDS0, KIDS0-m*87, KIDS-A, and KIDS-B models are depicted in Table~\ref{tab1}. Numerical values of the Skyrme force parameters and nuclear matter properties of the SLy4 model can be found in the Skyrme-Lyon model paper \cite{sly4}.

%%%TABLE 1 %%%%%%%
\begin{table}[t]
  \caption{ Skyrme force parameters for the KIDS-EDF models. 
The units of $t_0, y_0$ are in MeV fm$^{3}$, the units of $t_{31}, y_{31}$ are in MeV fm$^{4}$, the units of $t_1$, $t_2$, $t_{32}$, $y_{32}$, $W_0$ are in MeV fm$^{5}$, and the unit of $y_{33}$ is in MeV fm$^{6}$. $K_0$, $J$, and $L$ are in units of MeV. At the bottom, we summarize the data used in determining the model parameters. $R^{1.4}_{\rm NS}$, $E_{\rm B}/A$, and $R_c$ denote the radius in km of a neutron star with mass $1.4 M_\odot$, binding energy per nucleon and charge radius, respectively.}
  \label{tab1}
  \addtolength{\tabcolsep}{6.5pt}
  \begin{tabular}{ccccc} 
    \hline \hline
    Parameters & KIDS0 & KIDS0-m*87 & KIDS-A & KIDS-B  \\[0.2em] 
    \hline
    $t_0$   & $-1772.044$ & $-1772.044$ & $-1855.377$ & $-1772.044$ \\
    $y_0$   & $-127.524$  & $-127.524$ & 2182.404 & 2057.283  \\
    $t_1$   &  275.724  &  $376.694$ & 276.058 & 271.712 \\
    $y_1$   &  0.000    &  $-265.280$ & 0.000  & 0.000  \\
    $t_2$   & $-161.507$  & $145.531$ &$-167.415$ & $-161.957$ \\
    $y_2$   & 0.000        & $-334.837$ & 0.000 & 0.000  \\
    $t_{31}$ & 12216.730 & 12216.730 & 14058.746 & 12216.730  \\
    $y_{31}$ & $-11969.990$ & $-11969.990 $& $-73482.960$ & $-70716.593$  \\
    $t_{32}$ & 571.074 & $-1233.167$ & $ -1022.193$ & 622.750  \\
    $y_{32}$ & 29485.421 &  $32553.976$ & 122670.831 & 119258.903 \\
    $t_{33}$ & 0.000 & 0.000 & 0.000  & 0.000  \\
    $y_{33}$ &$ -22955.280$ &$ -22955.280$ & $ -73105.329 $&$ -70290.560$ \\
    $W_0$   & 108.359 & 133.722 & 92.023 &  91.527 \\ \hline
    $\mu^*_s$ & 0.991 & 0.800 & 1.004 & 0.997  \\
    $\mu^*_v$ & 0.819 & 0.700 & 0.827 & 0.825 \\ 
    $K_0$ & 240 & 240 & 230 & 240 \\
    $J$ & 32.8 & 32.8 & 33 & 32\\
    $L$ & 49.1 & 49.1 & 66 & 58 \\ \hline
Input data & APR PNM EoS & $R^{1.4}_{\rm NS} = $ 11.8-12.5 & $\leftarrow$ & $\leftarrow$ \\
in the & $E_{\rm B}/A$ and  $R_c$ of &  $\leftarrow$ & $\leftarrow$ & $\leftarrow$ \\
fitting & $^{40}$Ca, $^{48}$Ca, $^{208}$ Pb & & \\ 
    \hline \hline
  \end{tabular}
\end{table}

%------------SECTION 3 ----------------------
\section{Neutron star property}
\label{sec:MRrelation}
%---------------------------------------------

%%% FIG1 %%%%%%%%%
\begin{figure}[t]
  \begin{center}
    \includegraphics[width=0.48\textwidth]{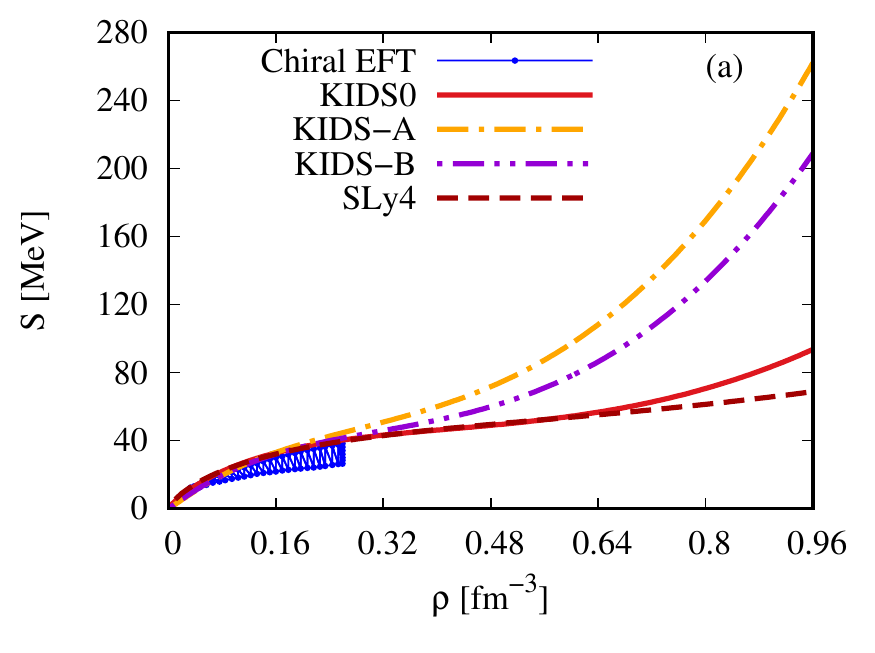}
    \includegraphics[width=0.48\textwidth]{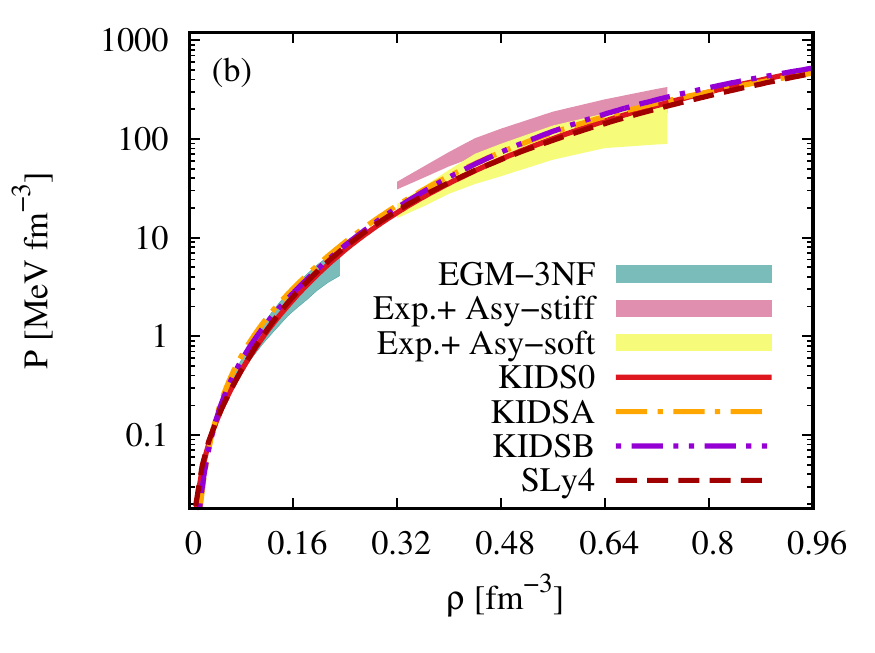}    
  \end{center}
  \caption{(a) Nuclear symmetry energy $\mathcal{S}(\rho)$ and (b) pressure for the PNM as a function of $\rho$ for the considered models.}
  \label{fig1}
\end{figure}

Neutron star properties are calculated from the EoS, which denotes the pressure as a function of the energy density. From the thermodynamic relation, pressure is obtained by differentiating the energy density with respect to density. Therefore, the energy density or energy per particle given by Eq.~(3) is the most fundamental quantity that 
determines the NS properties. In the core of NSs, expecting $\delta \simeq 1$, the behavior of symmetry energy is crucial in determining the NS EoS. Results for the nuclear symmetry energy $\mathcal{S}(\rho)$ of the considered models are shown as a function of density in Fig.~\ref{fig1}(a). It shows that KIDS0 and KIDS0-m*87 models have the same symmetry energy since both models are calibrated to identical nuclear matter properties. For this reason, the curve of symmetry energy for the KIDS0-m*87 model is omitted in Fig.~1(a). In the low densities up to $\rho \simeq \rho_0 =$ 0.16 fm$^{-3}$, all the models have quite similar behavior of the nuclear symmetry energies and they are consistent with the result from the chiral effective theory calculation at low densities~\cite{Sammarruca:2014zia}. Model dependence of the symmetry energy becomes evident at densities much higher than $\rho_0$. The stiffness of symmetry energies of the models is given in the order of KIDS-A $>$ KIDS-B $>$ KIDS0 $\simeq$ SLy4. Stiffness could be understood easily by noting the values of $L$. In Tab.~\ref{tab1}, we show that $L$(KIDS-A) $>$ $L$(KIDS-B) $>$ $L$(KIDS0), and $L$(SLy4)$=45.9$ MeV, which exactly gives the same ordering as the stiffness of symmetry energy at high densities.

In Fig.~\ref{fig1}(b), we show the results of the pressure for the PNM as a function of density. At low density, the pressures for all the models have excellent agreement with the predictions obtained from chiral perturbation theory (ChPT)~\cite{Tews:2012fj}. At high density, the results of all the models fit well with HIC data~\cite{Danielewicz:2002pu}.

Mass splitting of the neutron and proton in dense nuclear matter can provide valuable information on the charge symmetry breaking of nuclear interactions at finite density.
The effective mass of the nucleon given by Eq.~(\ref{eq:efmn}) is determined by the nuclear forces $\Theta_s$, $\Theta_v$, isospin of the nucleon $\tau^i_3$, and the asymmetry of the nuclear matter $\delta$. In the core of the neutron star, at a given baryon density $\rho$, $\rho_n$ and $\rho_p$ are determined by the $\beta$-equilibrium condition
\begin{eqnarray}
  \label{eq3a2}
  \mu_n =  \mu_p + \mu_e,
\end{eqnarray}
and the charge neutrality
\begin{eqnarray}
  \label{eq3a3}
  \rho_p = \rho_e + \rho_\mu,
\end{eqnarray}
where $\rho_e$ and $\rho_\mu$ are the densities of the electrons and the muons, respectively. Chemical potentials of the neutrons and the protons are calculated from the standard relation  $\mu_{(p,n)} = \frac{\partial \mathcal{E} (\rho, \delta)}{\partial \rho_{(p,n)}}$. Electrons and muons are approximated to be free, so their chemical potentials are given by $\mu_{(e,\mu)} = \sqrt{k_{F(e,\mu)}^2 + m_{(e,\mu)}^2}$, where $m_e$ and $m_\mu$ are the electron and muon masses in free space, respectively.We note that, in the present calculation, we neglect the interaction of neutrinos with electrons and muons since their contributions to the cross section are marginal in comparison with those of the protons and neutrons.

%%%% FIG2 %%%%%%%%%%%
\begin{figure}[t]
  \begin{center}
    \includegraphics[width=0.48\textwidth]{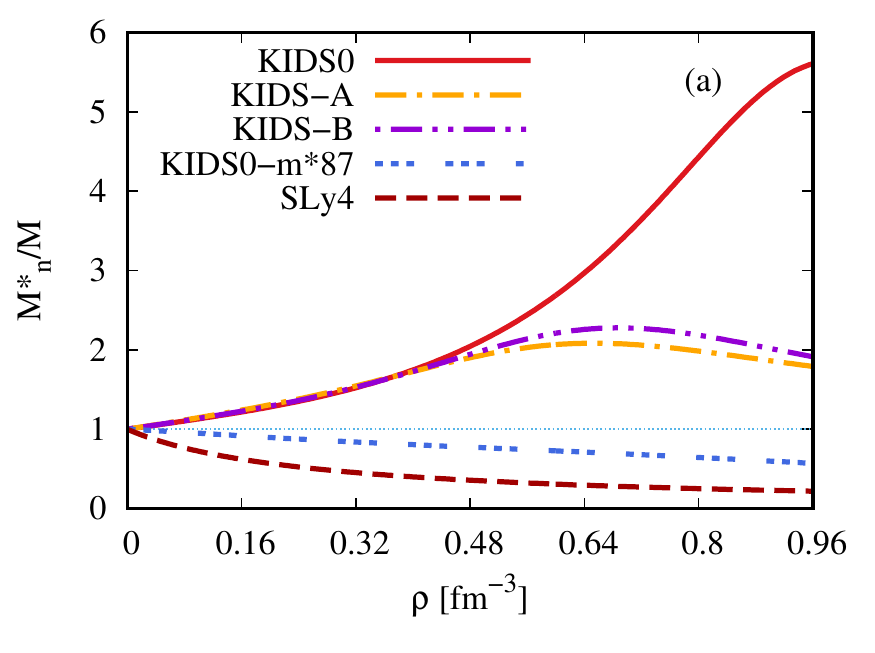}
    \includegraphics[width=0.48\textwidth]{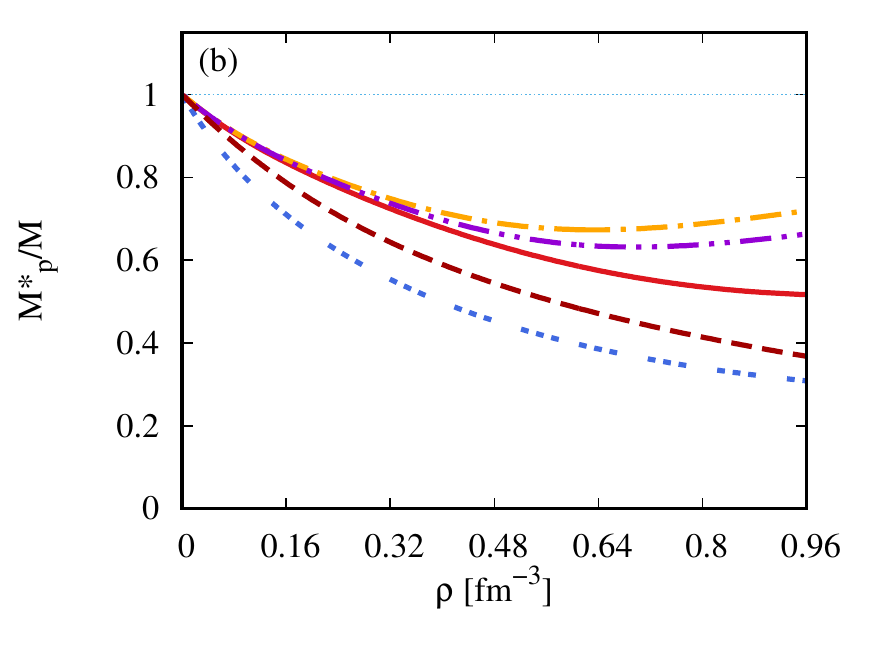}
  \end{center}
  \caption{Effective masses of the neutrons (a) and protons (b) for the considered models.}
  \label{fig2}
\end{figure}

Effective mass in the NS matter is shown for the neutron in Fig.~\ref{fig2}(a) and for the proton in Fig.~\ref{fig2}(b). The most notable result is the behavior of $M^*_n$: It is larger than the free mass in the KIDS0, KIDS-A, and KIDS-B models, but it is the opposite in the KIDS0-m*87 and SLy4 models. Neutron effective mass can be rewritten in terms of $\mu^*_s$ and $\mu^*_v$ as 
\begin{eqnarray}
M^*_n = M_n \left[ 1 + \frac{M_n}{8 \hbar^2} \rho \Theta_s - 
\left( \frac{1}{\mu^*_v} - \frac{1}{\mu^*_s} \right) \delta \right]^{-1}.
\label{eq:neutronefm}
\end{eqnarray}
In the KIDS0, KIDS-A, and KIDS-B models, $\mu^*_s$ is larger than $\mu^*_v$ by about 0.2, but the difference is 0.1 in the KIDS0-m*87 and SLy4 models. A large difference between $\mu^*_s$ and $\mu^*_v$ leads to a large negative contribution $-(1/\mu^*_v - 1/\mu^*_s)$ in the denominator of the neutron effective mass and this results in $M^*_n > M_n$. The effects of the isoscalar and isovector effective mass difference are furthermore tested by considering $\mu^*_s=0.9$ and $\mu^*_v=0.7$, leading to $M^*_n$ behaving similarly to the KIDS0 model. This result confirms that the neutron effective mass is sensitive to the difference between $\mu^*_s$ and $\mu^*_v$.

Neutron effective masses in the KIDS0, KIDS-A, and KIDS-B models are similar up to $\rho \simeq 0.48$~fm$^{-3}$. This similarity is understood well since they have similar $\mu^*_s$ and $\mu^*_v$ values. A sizable and vivid difference appears above $\rho = 0.48$~fm$^{-3}$. This difference mainly originates from the different results of the asymmetry $\delta$ in the KIDS0, KIDS-A, and KIDS-B models. If the symmetry energy is stiffer, more energy is needed to convert a proton to a neutron. Therefore, stiff symmetry energy favors less asymmetry, and the proton fraction $Y_p = \rho_p/\rho$ becomes larger than the soft symmetry energy. Figure \ref{fig3}(b) shows the particle fractions $Y_n=\rho_n/\rho$ and $Y_p=\rho_p/\rho$ within the neutron star. Neutron fraction is in the order $Y_n$(KIDS0) $>$ $Y_n$(KIDS-B) $>$ $Y_n$(KIDS-A), which is exactly the reverse of the inequality for the stiffness of symmetry energy. Sizable difference in $Y_n$ starts to appear around $\rho \simeq 0.32$~fm$^{-3}$ for the KIDS0, KIDS-A, and KIDS-B models. With smaller $\delta$ values in the KIDS-A and KIDS-B models than in the KIDS0 model, the effect of the negative contribution in the denominator of $M^*_n$ is suppressed. Consequently,  $M^*_n$ of the KIDS-A and KIDS-B models is smaller than that of the KIDS0 model.

In Fig.~\ref{fig2}(b), we show the result of $M^*_p/M$ for the KIDS0, KIDS0-m*87, KIDS-A, KIDS-B, and SLy4 models. In contrast to the neutron effective mass, all the models predict the proton effective mass smaller than the free state mass. This could be understood from the formula of $M^*_p$
\begin{eqnarray}
M^*_p = M_p \left[ 1 + \frac{M_p}{8 \hbar^2} \rho \left\{ 2 \delta \Theta_v +(1-\delta) \Theta_s \right\} \right]^{-1}.
\label{eq:protonefm}
\end{eqnarray}
Both $\Theta_s$ and $\Theta_v$ are positives, and $\delta \leq 1$ within the neutron star. Therefore, the term added to 1 in the denominator is always positive, so the denominator is always larger than 1, and it gives $M^*_p$ smaller than $M_p$. Results of the KIDS0, KIDS-A, and KIDS-B models are similar up to $\rho \simeq 0.48$~fm$^{-3}$ and they deviate from each other at higher densities. Within the same models, such behavior is also observed in $M^*_n$. Those behaviors can be explained well in terms of the different results of $\delta$ from each model.

The dependence of $M^*_p$ on $\mu^*_s$ and $\mu^*_v$ can be understood well  by assuming $\delta \simeq 1$, which is a reasonable approximation up to $\rho \simeq 2 \rho_0$ as in Fig.~\ref{fig3}(b). With $\delta=1$, $M^*_p$ becomes equal to the isovector effective mass $m^*_v$. The smallest isovector effective mass is found in the KIDS-m*87 model and it is a quite similar value in the KIDS0, KIDS-A, and KIDS-B models. The magnitude of $M^*_p$ is arranged exactly in agreement with the order of $\mu^*_v$. A summary of the typical effective neutron masses and the nuclear symmetry energies for the KID-EDF models is given in Table~\ref{tab2}.

%%TABLE 2%%%%%%%%%%
\begin{table}[t]
  \caption{Effective neutron  masses and nuclear symmetry energies for the
    KIDS-EDF models.}
  \label{tab2}
  \addtolength{\tabcolsep}{16.6pt}
  \begin{tabular}{cc} 
    \hline \hline
    Type of KIDS-EDF model & $M_n^*/M$ and feature of the symmetry energy  \\[0.2em]\hline
    KIDS0 & $M_n^* /M \gtrsim 1$ and soft symmetry energy  \\
    KIDS0-m*87 & $M_n^* /M \lesssim 1$ and soft symmetry energy \\
    KIDS-A & $M_n^* /M \gtrsim 1$ and stiff symmetry energy \\
    KIDS-B & $M_n^* /M \gtrsim 1$ and stiff symmetry energy \\ 
    SLy4  &  $M_n^* /M \lesssim 1$ and soft symmetry energy \\ 
    \hline \hline
  \end{tabular}
\end{table}

The mass and radius of a neutron star can be determined by solving the TOV equations for the static (non rotating) NS, which are respectively given by~\cite{Tolman:1939jz,Oppenheimer:1939ne}
\begin{eqnarray}
  \label{eq4}
  \frac{dP(r)}{dr} &=& - \frac{G[M(r) + 4 \pi r^3 P(r)/c^2][\mathcal{E}(r) + P(r)]}{r[r-2GM(r) /c^2] c^2}, \nonumber \\
  \frac{dM(r)}{dr} &=& 4\pi r^2 \frac{\mathcal{E}(r)}{c^2},
\end{eqnarray}
where the radial distance from the center is symbolized by $r$ and $M(r)$ is the mass profile of the neutron star within $r$. $P(r)$ and $\mathcal{E}(r)$ are respectively pressure and energy density obtained from the nuclear EDF models. Both equations in~(\ref{eq4}) are numerically solved using the Runge-Kutta integration technique by integrating them over the radial distance from the center up to the surfaces of the NS where $P(R_{\textrm{NS}}) = 0$, from which radius of the star $R_{\rm NS}$ is determined, and the star mass is obtained from $M_{\rm NS} = M(r=R_{\rm NS})$.

%%%FIG3%%%%%%
\begin{figure}[t]
  \begin{center}
    \includegraphics[width=0.48\textwidth]{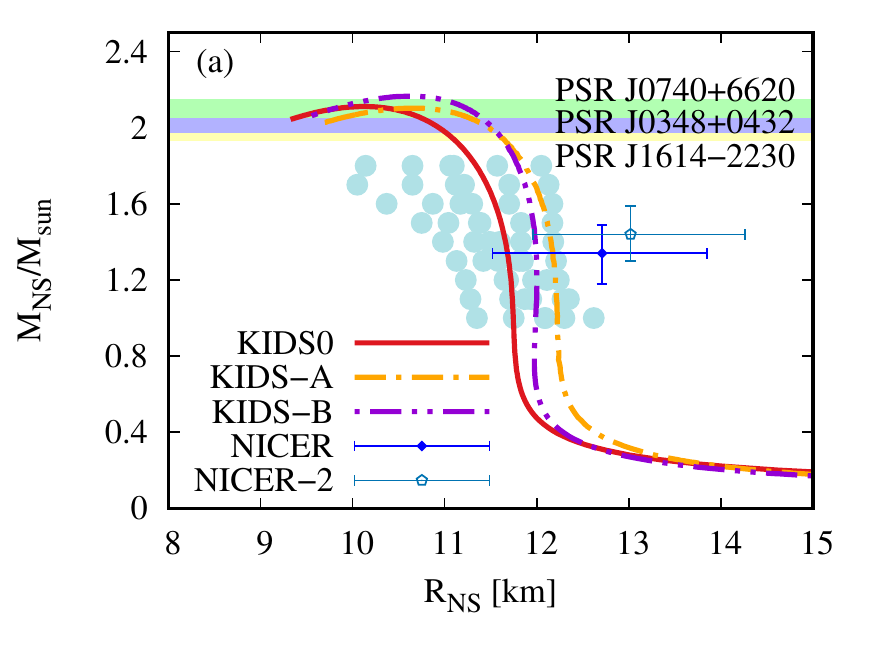}
    \includegraphics[width=0.48\textwidth]{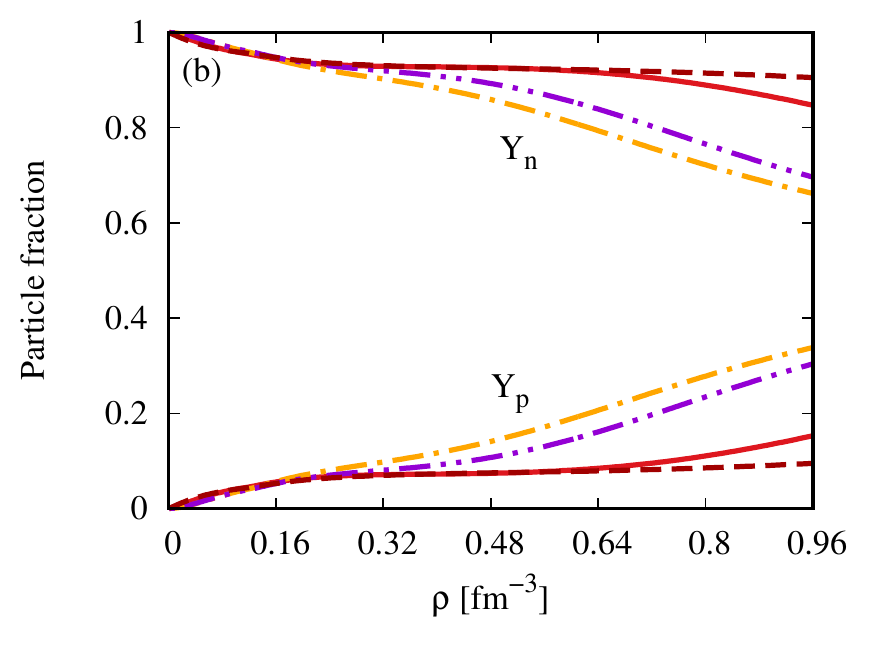}
  \end{center}
  \caption{(a) Neutron star mass-radius relations and (b) particle fractions for neutrons and protons for the considered models.}
  \label{fig3}
\end{figure}

Results for the NS mass as a function of the NS radius $R_{\textrm{NS}}$ and particle fractions for neutrons and protons as a function of density are shown in Fig.~\ref{fig3}. Figure~\ref{fig3}(a) shows that the prediction results for $M_\mathrm{NS}/M_{\textrm{sun}}$ with the KIDS0 and SLy4 models are quite similar, where these models have similar soft symmetry energies, as already shown in Fig.~\ref{fig1}(a). In comparison with the results of the KIDS0 model,
the KIDS-A and KIDS-B models with stiff symmetry energies predict a larger radius of the NS. Stiff EoS implies that the matter is less compressible and as a result, the size of the system becomes larger. As already shown in Fig.~\ref{fig1}(a), the symmetry energy is stiff in the order of KIDS-A $>$ KIDS-B $>$ KIDS0 $\simeq$ SLy4. The result for $R_{\rm NS}$ in the NS mass range $(0.8-1.6)M_{\rm sun}$ shows exactly the same ordering as the stiffness of symmetry energy. This obviously demonstrates that the symmetry energy plays a crucial role in the bulk properties of NS. On the other hand, the NS mass and radius results for all the KIDS-EDF models are consistent with the recent observations of PSR J0348+0432~\cite{Antoniadis:2013pzd}, PSR J0740+6620~\cite{NANOGrav:2019jur}, and PSR J1614-2230~\cite{Demorest:2010bx}, 
predicting the maximum mass of NS larger than 2.0 $M_\mathrm{sun}$, as well as the NICER results for radius $R_{\textrm{NS}} =$ 12.35 $\pm$ 0.75 km with a NS mass $M_{\textrm{NS}} =$ 2.08 $M_\mathrm{sun}$~\cite{Miller:2021qha}. The blue dots denote the M-R range obtained from the analysis of low mass x-ray binary (LMXB) data~\cite{Steiner:2010fz}.

In Fig.~\ref{fig3}(b) we show the results of the particle fraction for the neutron and proton. At densities up to $\rho \simeq 2 \rho_0$, particle fractions are similar in the considered models. The difference between the models becomes obvious as the density increases. At $\rho \gtrsim 3 \rho_0$, based on the size magnitude of the particle fraction, the results can be classified into two groups: one group with KIDS0, SLy4 and the other with KIDS-A, KIDS-B. The former group keeps a large neutron fraction at high densities. If the symmetry is soft, the energy cost to convert a proton to a neutron is less than the one for the stiff symmetry energy. As a consequence of the soft symmetry energy, the system can be in the ground state with more neutrons than the stiff symmetry energy. Grouping of the particle fraction agrees well with the stiffness of the symmetry energy of the models. The small difference between the KIDS-A and KIDS-B models is also attributed to the stiffer symmetry energy of KIDS-A.

%--------------------------------------------------
\section{Neutrino Mean Free Path} \label{sec:NMFP}
%--------------------------------------------------  
%

In the section, we present the neutrino interaction with constituents of NS matter. The Lagrangian density of neutrino interactions with NS constituents matter via current-current interaction is defined by~\cite{Hutauruk:2021cgi,Reddy:1997yr,Hutauruk:2010tn,Niembro:2001hd}
\begin{eqnarray}
  \label{eq5}
  \mathscr{L}_{\textrm{int}}^{(n,p)} &=& \frac{G_F}{\sqrt{2}} \left[ \bar{\nu}_e 
\gamma^\mu (1 - \gamma_5) \nu_e \right] \left[\bar{\psi} \Gamma_\mu^{(n,p)} \psi \right],
\end{eqnarray}
where the nucleon vertex is defined by $\Gamma_\mu^{(n,p)} = \gamma_\mu ( C_V^{(n,p)} - C_A^{(n,p)} \gamma_5 )$ and the weak coupling constant is $G_F = 1.023 \times 10^{-5}/M^2$. The vector and axial coupling constants are $C_V =-0.5$ and $C_A = -g_A /2$ for the neutron, whereas $C_V = 0.5 - 2 \sin^2 \theta_w$ and $C_A = g_A/2$ for the proton,  where $g_A = 1.260$ and $\sin^2 \theta_w = 0.223$. Note that the Lagrangian density for the charged-current absorption reaction is the same as the neutral-current scattering, which leads to the same expressions of the DCRS. Only the values of the axial and vector coupling constants are different~\cite{Reddy:1997yr}.

From the Lagrangian in Eq.~(\ref{eq5}) we easily derive the DCRS of the neutrino and one has
\begin{eqnarray}
  \label{eq6}
  \frac{1}{V} \frac{d^3 \sigma}{dE_\nu' \,d^2\Omega} &=& - \frac{G_F^2}{32\pi^2} \frac{ E_\nu'}{E_\nu} \Im \left[ L_{\mu \nu} \Pi^{\mu \nu} \right],
\end{eqnarray}
where $E_\nu$ and $E_\nu'$ are respectively the initial and final neutrino energies. The polarization tensors $\Pi^{\mu \nu}$ for the target neutrons and protons are given by
\begin{eqnarray}
  \label{eq7}
  \Pi_{\mu \nu}^{(n,p)} (q^2) &=& -i \int \frac{d^4p}{(2\pi)^4} \textrm{Tr} \left[ G^{(n,p)} (p) \Gamma_{\mu}^{(n,p)} G^{(n,p)} (p + q) \Gamma_\nu^{(n,p)} \right],
\end{eqnarray}
where $G^{(n,p)}$ are the propagators and $p =(p_0,\mathbf{p})$ is the initial four-momentum of the neutron and proton targets. The propagators of neutrons and protons are explicitly expressed by
\begin{eqnarray}
  \label{eq8}
  G^{(n,p)} (p) &=& \left[ \frac{p\!\!\!/^* + M^*}{p^{*2} -M^{*2} + i \epsilon} + i \pi \frac{p\!\!\!/^* + M^*}{E^*} \delta \left( p_0^* - E^* \right) \Theta \left( p_F^{(n,p)} - |\mathbf{p}|\right) \right],
\end{eqnarray}
where $E^* = E + \Sigma_0 = \sqrt{\mathbf{p}^{*2} + M^{*2}}$ is the effective nucleon energy, and $M^* = M + \Sigma_s$ represents the nucleon effective mass. $\Sigma_s$ and $\Sigma_0$ are the scalar and time like self-energies, respectively. $\mathbf{p}^* = \mathbf{p} + \left( \frac{\mathbf{p}}{|\mathbf{p}|} \right) \Sigma_v$ is the nucleon effective momentum with $\mathbf{p}$ and $\Sigma_v$ being the three-component momentum of a nucleon and the space like self-energy, respectively. 
$p^{(n,p)}_{F} = \sqrt{E^{(n,p)2}_{F} - M^{*(n,p)2}}$ are the proton and neutron Fermi momenta. Note that the $\Sigma_s$, $\Sigma_0$, and $\Sigma_v$ are obtained by solving the Schr\"odinger-equivalent single-nucleon potential $\mathcal{V} (E)$ in relativistic models, which contains the central and spin-orbit potentials as in Ref.~\cite{Jaminon:1989wj} and references therein. The central potential part, which is called the Schr\"odinger-equivalent optical potential, has the form $\mathcal{V}_c (E) = \Sigma_s - \left( E/M + 1\right) \Sigma_0 + (\Sigma_s^2 - \Sigma_0^2)/2M$ with the energy-momentum relation: $E = p^2/2M + \mathcal{V}_c (E)$. Using the dispersion relation, one finally gets $M^* = M  + \Sigma_s$. In the KIDS-EDF models, both potentials are written in terms of the standard density-dependent central and spin-orbit Skryme force parameters as in Eq.~(\ref{eq2a})~\cite{Vautherin:1971aw} and we then obtain the nucleon effective mass $M_i^*$ of Eq.~(\ref{eq:efmn}). The neutrino tensor $L_{\mu \nu}$ is then given by
\begin{eqnarray}
  \label{eq9}
  L_{\mu \nu} &=& 8 \left[ 2 k_\mu k_\nu + (k\cdot q) g_{\mu\nu} - (k_\mu q_\nu + q_\mu k_\nu) - i \epsilon_{\mu \nu \alpha \beta} k^\alpha q^\beta \right],
\end{eqnarray}
where $q = (q_0,\mathbf{q})$ is the four-momentum transfer and $k =(k_0, \mathbf{k})$ stands for the initial neutrino four-momentum.

After contracting the neutrino tensor in Eq.~(\ref{eq9}) and the polarizations for the neutrons and protons given in Eq.~(\ref{eq7}), 
the final expression for the neutrino DCRS is given by
\begin{eqnarray}
  \label{eq9a}
  \frac{1}{V} \frac{d^3 \sigma}{dE_\nu' \,d^2\Omega} &=& \frac{G_F^2}{4\pi^3} 
\frac{E_\nu^{'}}{E_\nu} q^2 \left[ A \mathscr{R}_1 + \mathscr{R}_2 + B \mathscr{R}_3\right],
\end{eqnarray}
where $A = [2E_\nu(E_\nu -q_0) + 0.5\,q^2]/|\mathbf{q}|^2$, $B = 2E_\nu -q_0$, and 
$\mathscr{R}_1$, $\mathscr{R}_2$, and $\mathscr{R}_3$ are respectively defined as
\begin{eqnarray}
  \label{eq9b}
  \mathscr{R}_1 &=& ( C_V^2 + C_A^2 ) [\Im \Pi_L^{(p,n)} + \Im \Pi_T^{(p,n)}], \nonumber \\
  \mathscr{R}_2 &=& C_V^2 \Pi_T^{(p,n)} + C_A^2 [\Im \Pi_T^{(p,n)} - \Im \Pi_A^{(p,n)}], \nonumber \\
  \mathscr{R}_3 &=& 2 C_V C_A \Im \Pi_{VA}^{(p,n)}.
\end{eqnarray}
The polarization insertion for the longitudinal, transversal, axial, and mixed vector-axial channels of the protons and neutrons is given by
\begin{eqnarray}
  \label{eq9c}
  \Im \Pi_L &=& \frac{q^2}{2\pi|\mathbf{q}|^3} 
\left[\frac{q^2}{4}(E_F -E^{*}) + \frac{q_0}{2}(E_F^2 -E^{*2}) + \frac{1}{3}(E_F^3 - E^{*3}) \right], \nonumber \\
  \Im \Pi_T &=& \frac{1}{4\pi|\mathbf{q}|} \left[ (M^{*2} + \frac{q^4}{4|\mathbf{q}|^2} + \frac{q^2}{2}) (E_F - E^{*}) +\frac{q_0 q^2}{2|\mathbf{q}|^2} (E_F^2 - E^{*2}) + \frac{q^2}{3|\mathbf{q}|^2} (E_F^3 - E^{*3}) \right],\nonumber \\
  \Im \Pi_A &=& \frac{i}{2\pi |\mathbf{q}|} M^{*2} (E_F -E^{*}), \nonumber \\
  \Im \Pi_{VA} &=& \frac{iq^2}{8\pi|\mathbf{q}|^3} \left[ (E_F^2 - E^{*2}) + q_0 (E_F-E^{*})\right].
\end{eqnarray}

Finally, we present the neutrino mean free path for the neutral-current neutrino scattering. Considering a fixed baryon density, the inverse of the neutrino mean free path (opacity) is straightforwardly obtained by integrating the differential cross section in Eq.~(\ref{eq6}) over the three-component momentum transfer $|\mathbf{q}|$ and energy transfer $q_0$, which gives
\begin{eqnarray}
  \label{eq10}
  \lambda^{-1} (E_\nu) &=& 2 \pi \int_{q_0}^{(2 E_\nu - q_0)} d|\mathbf{q}| \int_0^{2E_\nu} dq_0 \frac{|\mathbf{q}|}{E_\nu E_\nu'}  \left[ \frac{1}{V} \frac{d^3 \sigma}{dE_\nu' d^2\Omega} \right],
\end{eqnarray}
where $E_\nu' = E_\nu -q_0$.

%%%%FIG4%%%%%
\begin{figure}[t]
  \begin{center}
    \includegraphics[width=0.48\textwidth]{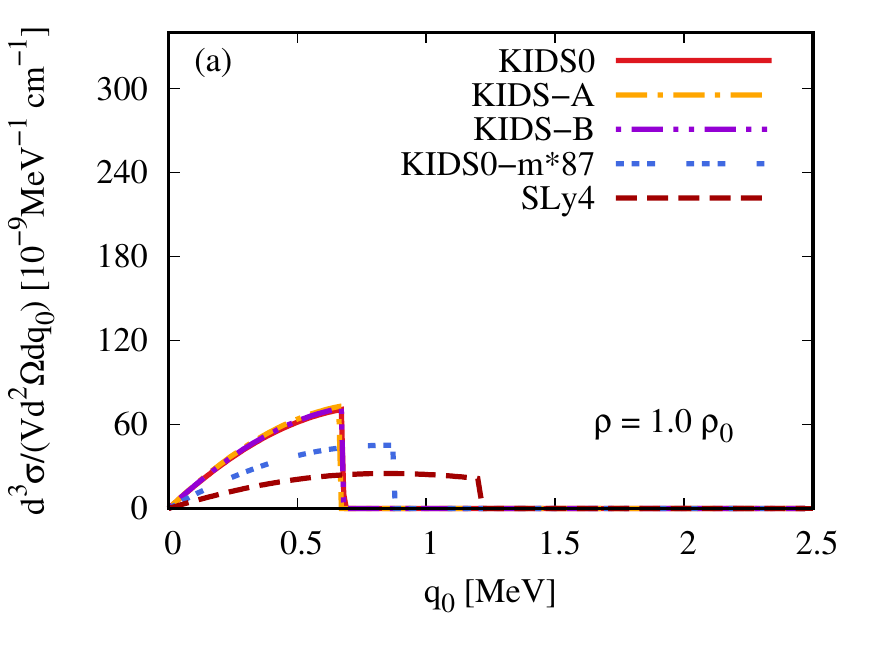}
    \includegraphics[width=0.48\textwidth]{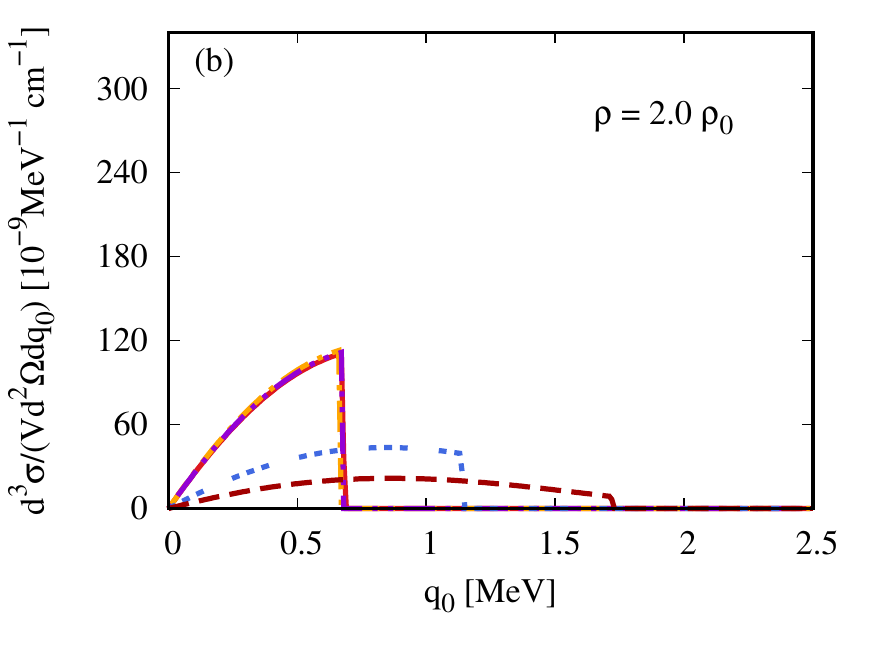}\\
    \includegraphics[width=0.48\textwidth]{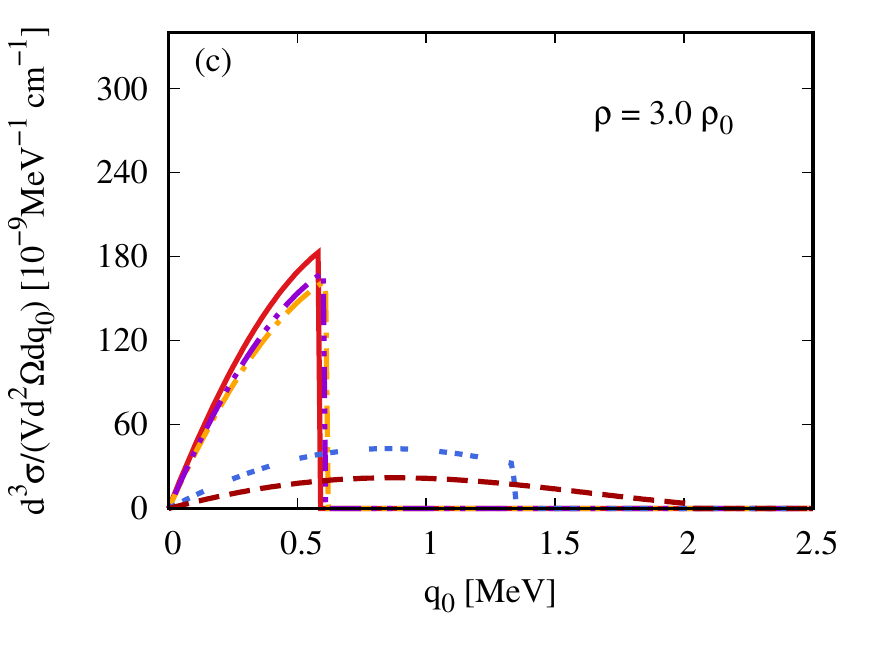}
    \includegraphics[width=0.48\textwidth]{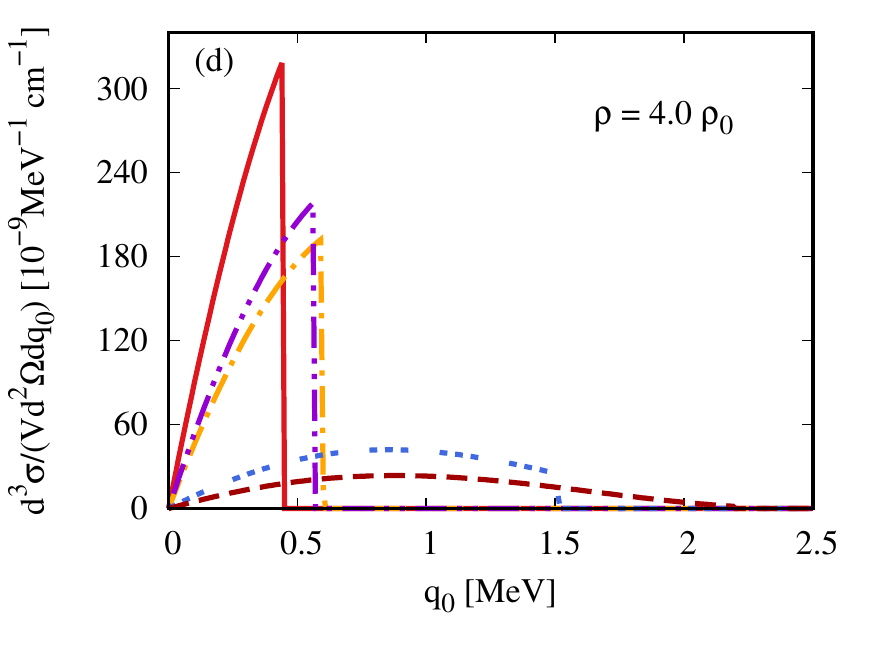}
  \end{center}
  \caption{DCRS of the neutrino with neutrons for various KIDS-EDF models as a function of $q_0$ at fixed $E_\nu =$ 5 MeV and $|\mathbf{q}| =$ 2.5 MeV for (a) $\rho = 1.0~\rho_0$, (b) $\rho = 2.0~\rho_0$, (c) $\rho = 3.0~\rho_0$ and (d) $\rho = 4.0~\rho_0$.}
  \label{fig4}
\end{figure}

Results for the DCRS of the neutrino off neutrons for the EDF models at fixed $E_\nu =$ 5 MeV, which is a typical kinematic for the NS cooling phase, and $|\mathbf{q}| =$ 2.5 MeV are shown in Fig.~\ref{fig4}. Figure~\ref{fig4}(a) shows that the magnitudes of the DCRS of the neutrino for the KIDS0, KIDS-A, and KIDS-B models with $M^*/M \gtrsim 1$ at $\rho =$ 1.0 $\rho_0$ are bigger than those obtained from the KIDS0-m*87 and SLy4 models with $M^* /M \lesssim 1$. However, the kinematic ranges of $q_0$ for the KIDS0, KIDS-A, and KIDS-B models are smaller than those obtained for the KIDS0-m*87 and SLy4 models. It is worth noting the relevance between the neutrino-neutron DCRS and $M^*_n/M$. As shown in Fig.~\ref{fig2}(a), the effective masses of neutrons for the KIDS0, KIDS-A, and KIDS-B models are rather similar up to $\rho \simeq 2.0~\rho_0$, which then leads to similar prediction results for the DCRS of the neutrino at the corresponding densities.

Similarities of the DCRS results amongst the KIDS0, KIDS-A, and KIDS-B models are kept at $\rho=2\rho_0$ as can be seen in Fig.~\ref{fig4}(b). The magnitude of the DCRS of the neutrino for the KIDS0 model is larger than those for the KIDS-A and KIDS-B models as the effective masses of neutrons become larger in the high-density regime ($\rho \gtrsim 2.0~\rho_0$). However, the magnitudes of the DCRS of the neutrino for the KIDS-A and KIDS-B models hold the same as their effective masses of neutrons are quite similar in the overall density regimes, which can be seen in Fig.~\ref{fig2}(a). Amongst the KIDS0, KIDS-A, and KIDS-B models, the different magnitudes of the DCRS of the neutrino 
are much more pronounced at $\rho = 4.0\rho_0$ as seen in Fig.~\ref{fig4}(d), where the effective mass of neutrons of the KIDS0 model is higher than that for the KIDS-A and KIDS-B models. Note that the DCRS of the neutrino drops abruptly to zero because the maximum value of the transfer energy $q_0^{\textrm{max}} \simeq |\mathbf{q}|/\sqrt{(M^*/p_F)^2 + 1}$ increases (decreases) as the effective nucleon masses decrease (increase) at a given value of the density or Fermi momentum $p_F$. The increase of the DCRS of the neutrino with larger $M^*_n/M$ values can be understood by the polarization insertions of Eq.(\ref{eq9c}).  Amongst the polarization insertions in Eq.~(\ref{eq9c}), the axial channel gives the largest contribution to the DCRS of the neutrino-neutrons scattering. With an approximation, $E_F - E^* \simeq \frac{1}{2 M^*}(p^2_F - p^{*2})$, Im$\Pi_A$ becomes proportional to $M^*$, and thus a large nucleon effective mass will increase the polarization insertion, leading to increasing DCRS of the neutrino. Analogously to the DCRS of the neutrino-neutrons scattering, a similar explanation can be given for the DCRS of neutrino-protons scattering as well as the DCRS of total scattering.

%%%%FIG5%%%%%%
\begin{figure}[t]
  \begin{center}
    \includegraphics[width=0.48\textwidth]{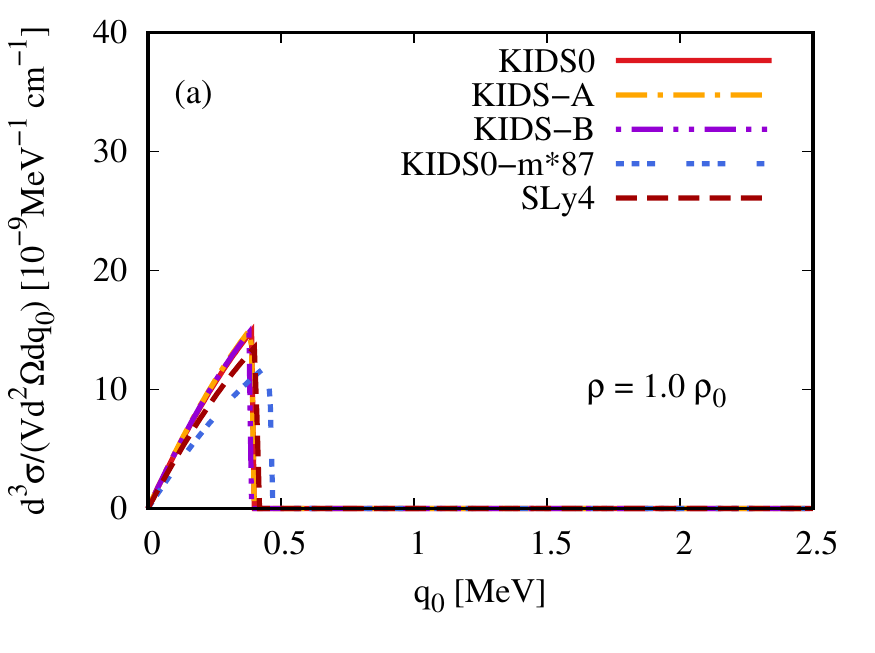}
    \includegraphics[width=0.48\textwidth]{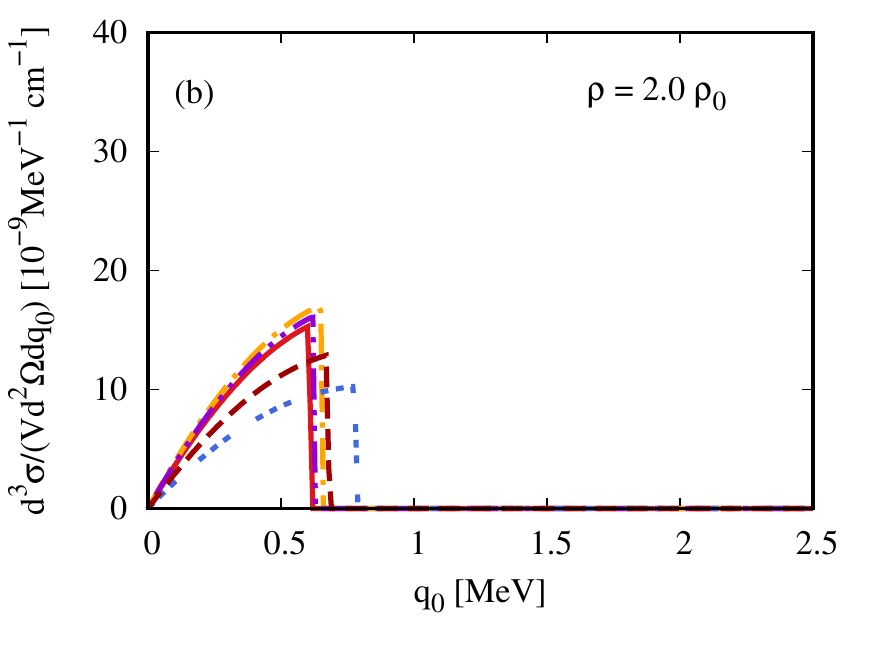}\\
    \includegraphics[width=0.48\textwidth]{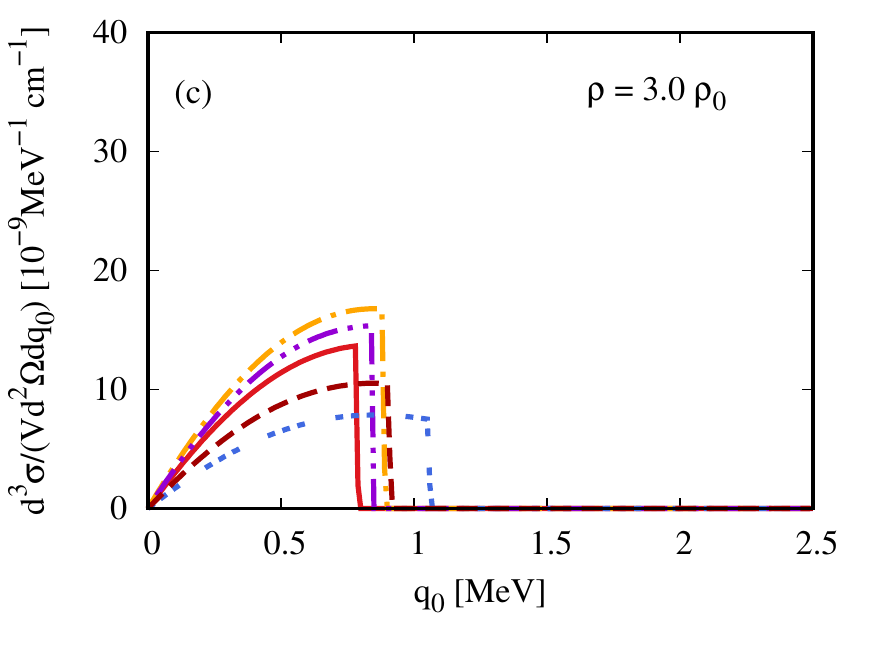}
    \includegraphics[width=0.48\textwidth]{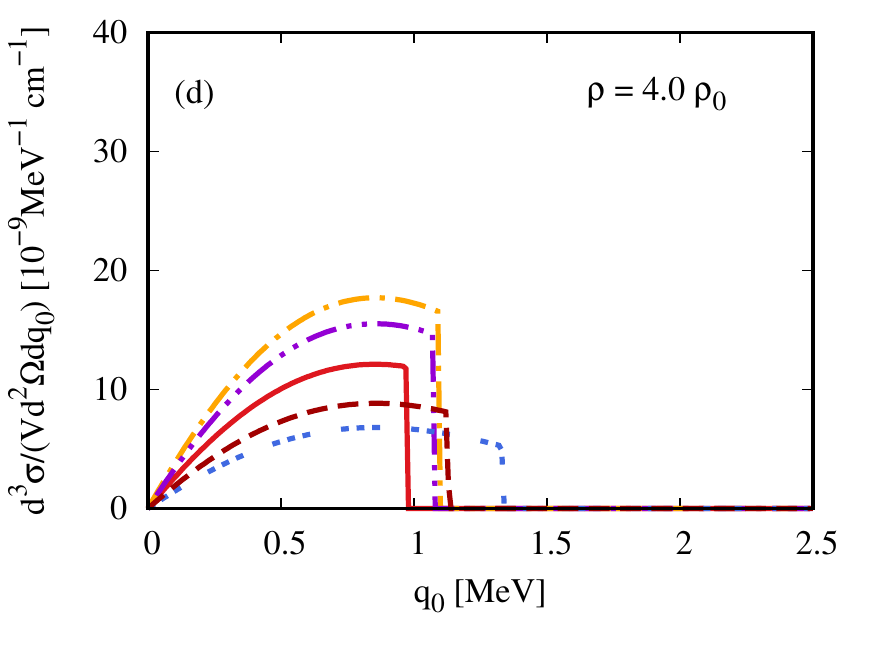}
  \end{center}
  \caption{Same as Fig.~\ref{fig4} but for 
    DCRS of the neutrino with protons.}
  \label{fig5}
\end{figure}

Figure~\ref{fig5} shows the DCRS of the neutrino with protons for the nuclear EDF models at fixed $E_\nu =$ 5 MeV and  $|\mathbf{q}| =$ 2.5 MeV as a function of $q_0$. 
At $\rho = 1.0~\rho_0$, the magnitude of the DCRS of the neutrino with protons is quite similar for the KIDS0, KIDS-A, KIDS-B, and SLy4 models, but it is rather different for the KIDS0-m*87 model, which has the lowest $M_p^* /M$. This behavior holds not only at $\rho = 1.0~\rho_0$ but also at higher densities as shown in Figs.~\ref{fig5}(b)-(d). The result confirms that, as shown in the scattering with neutrons, the DCRS of the neutrino is enhanced with a larger effective mass of the nucleon. The ranges of $q_0$ increase for all the models as the nuclear density increases. The largest range of $q_0$ is given by the KIDS0-m*87 model with the lowest $M_p^* /M$, as shown in Fig.~\ref{fig5}(d).

%%%%%FIG6%%%%%%%
\begin{figure}[t]
  \begin{center}
    \includegraphics[width=0.48\textwidth]{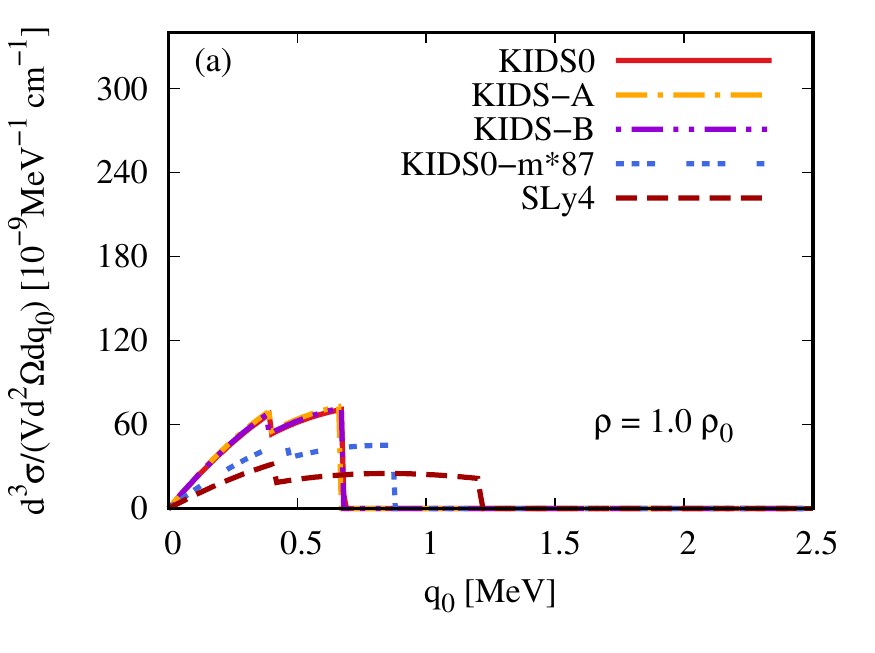}
    \includegraphics[width=0.48\textwidth]{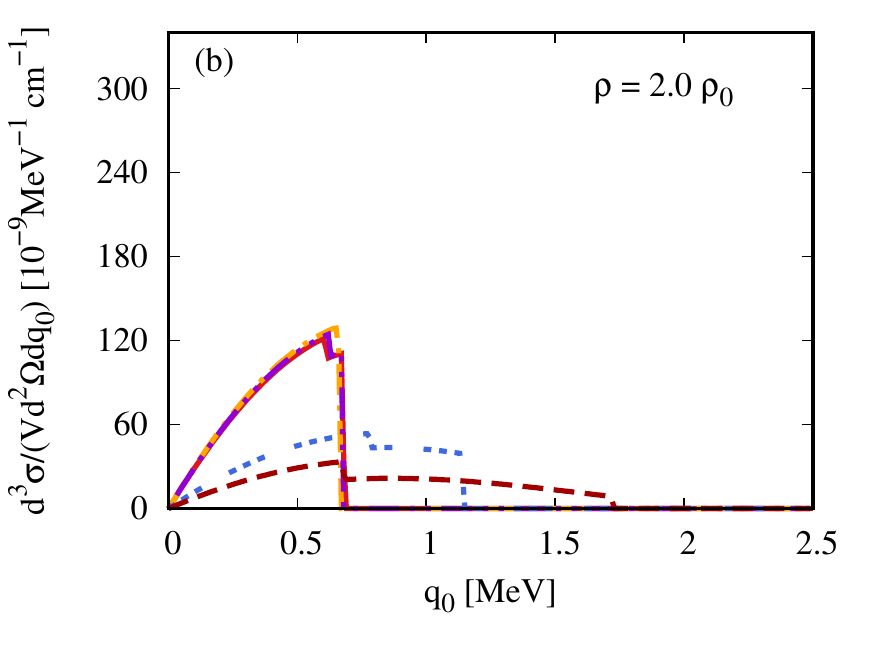}\\
    \includegraphics[width=0.48\textwidth]{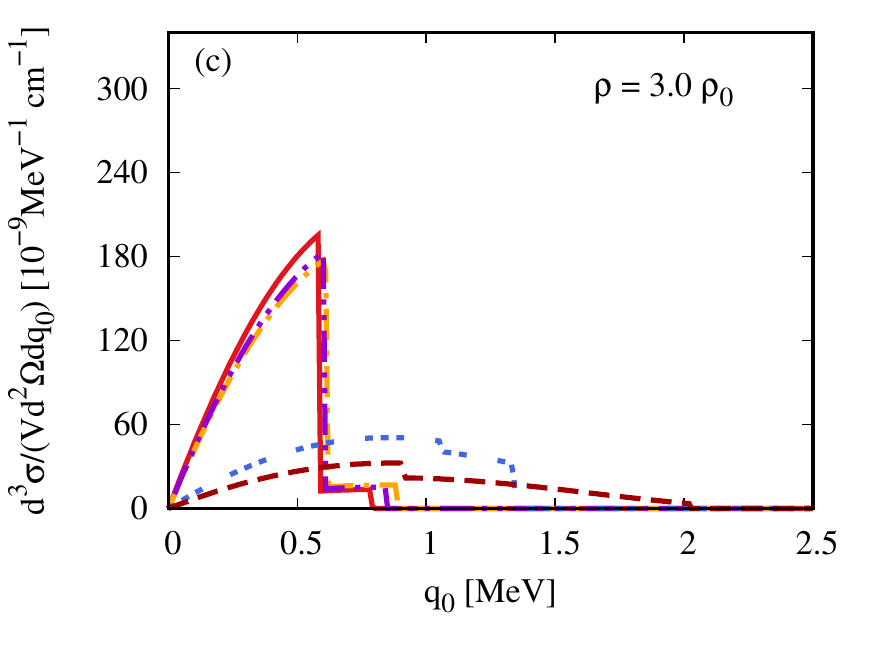}
    \includegraphics[width=0.48\textwidth]{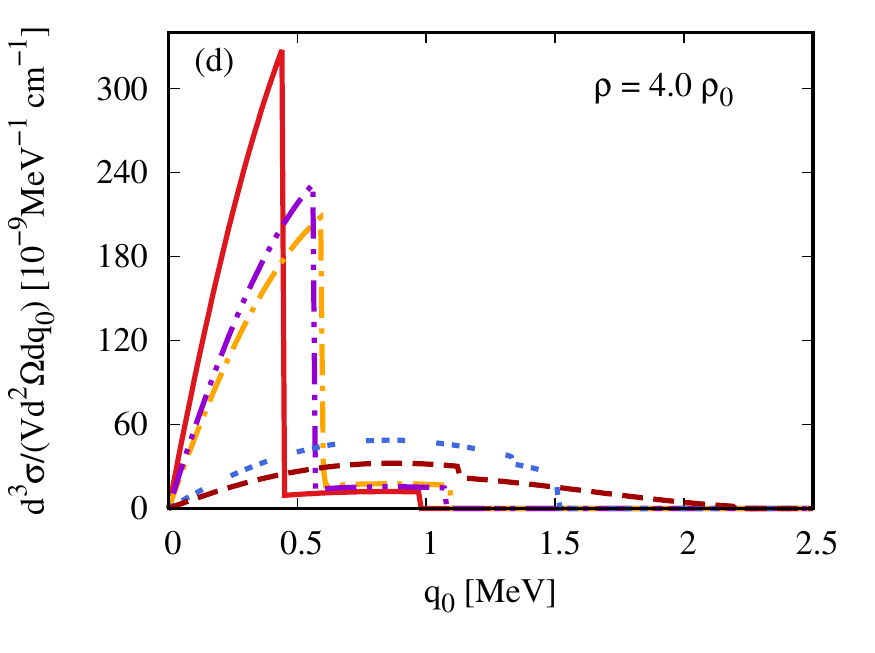}
  \end{center}
  \caption{Same as in Fig.~\ref{fig5} but for total 
    DCRS of the neutrino.}
  \label{fig6}
\end{figure}

The results for the total DCRS of the neutrino, which are obtained by summing up the DCRSs of the neutrino with protons in Fig.~\ref{fig5} and with neutrons in Fig.~\ref{fig4}, as a function of $q_0$ at fixed $E_\nu =$ 5 MeV and  $|\mathbf{q}| =$ 2.5 MeV are depicted in Fig.~\ref{fig6}. Figure~\ref{fig6} clearly shows the dominant contribution to the magnitude size of the total DCRS of the neutrino given by the DCRS of the neutrino with neutrons for the models with $M^*_n/M > 1$. However, the shape of the total DCRS of the neutrino depends on the shapes of DCRSs of the neutrino with both protons and neutrons. The total cross sections for the KIDS0, KIDS-A, and KIDS-B models increase as the density increases, but for the KIDS0-m*87 and SLy4 models the total cross sections of the neutrino are almost the same even if the density increases. Among the KIDS-EDF models, the biggest magnitude of the total cross section of the neutrino is given by the KIDS0 model and it is more pronounced at higher density. Consequently, the NMFP of the KIDS0 model is smaller than other KIDS-EDF models, as seen in Fig.~\ref{fig7}.

%%%%FIG7%%%%%%%%
\begin{figure}[t]
  \begin{center}
    \includegraphics[width=0.60\textwidth]{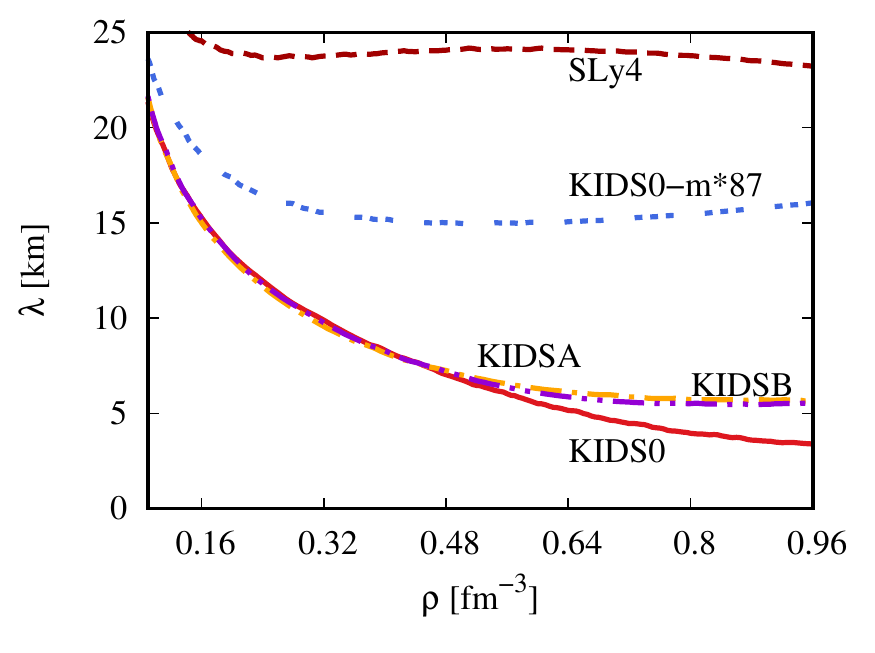}
  \end{center}
  \caption{Neutrino mean free path for various KIDS-EDF models as a function of $\rho$ at $E_\nu =$ 5 MeV and $|\mathbf{q}| =$ 2.5 MeV.}
  \label{fig7}
\end{figure}

Results for the NMFP for the five models at fixed $E_\nu =$ 5 MeV and $|\mathbf{q}| =$ 2.5 MeV as a function of the nuclear density are shown in Fig.~\ref{fig7}. NMPFs for the KIDS-A and KIDS-B models are quite similar for all regimes of the baryon densities. This behavior is followed by the KIDS0 model up to $\rho \simeq 3.0~\rho_0$ and then the NMFP for the KIDS0 model decreases faster than those obtained for the KIDS-A and KIDS-B models. Remarkable results on the NMFPs are predicted by the KIDS0-m*87 and SLy4 models. The NMFP predictions for the KIDS0-m*87 and SLy4 models are significantly higher compared with those obtained for the KIDS0, KIDS-A, and KIDS-B models. 
The magnitude of the NMPF for the SLy4 model is much bigger than that for the KIDS-EDF models.

%%%%%FIG8%%%%%
\begin{figure}[t]
  \begin{center}
    \includegraphics[width=0.65\textwidth]{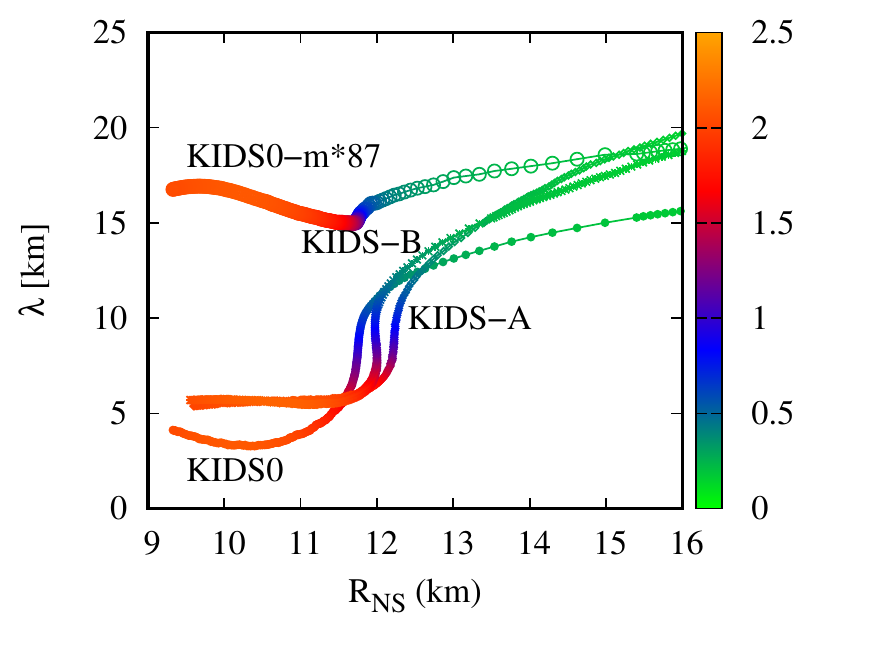}
  \end{center}
  \caption{
 Neutrino mean free path calculated at fixed $E_\nu =$ 5 MeV and $|\mathbf{q}| =$ 2.5 MeV, for various KIDS-EDF models as a function of NS radius $R_\mathrm{NS}$ and $M_{\textrm{NS}}/M_{\textrm{sun}}$.}
  \label{fig8}
\end{figure}

Results for the NMFP at fixed $E_\nu =$ 5 MeV and $|\mathbf{q}| =$ 2.5 MeV as a function of the NS radius for the considered models are shown in Fig.~\ref{fig8}. For the SLy4 model, for example, if we take $R_{\textrm{NS}} = 10$ km, it predicts $\lambda \simeq$ 22 km with NS mass $M_\mathrm{NS}/M_\mathrm{sun} \simeq$ 2.0, which indicates that the neutrino can escape from the NS, not being hindered by scattering with nucleons or the reabsorption process. It then leads to NS cooling faster than the case in which $R_{\textrm{NS}} > \lambda$. Similarly, for the KIDS0-m*87 model with $R_{\textrm{NS}} =$ 10 km, one predicts $\lambda \simeq $ 17 km with $M_\mathrm{NS} / M_\mathrm{sun} \simeq $ 2.0, showing that the neutrino can escape freely from the NS.

The KIDS0, KIDS-A and KIDS-B models with a similar $R_{\textrm{NS}} =$ 10 km predict respectively $\lambda \simeq$ 3.5 km with $M_{\textrm{NS}} /M_\mathrm{sun} \simeq$ 2.0 and $\lambda \simeq$ 5.03 and 5.04 km with $M_{\textrm{NS}} /M_\mathrm{sun} \simeq$ 2.0. The neutrinos will interact with the NS matter, leading to a NS cooling process slower than the KIDS-m*87 and SLy4 models because $R_{\textrm{NS}} > \lambda$.

Now let us take the canonical mass of neutron stars $M_{\rm NS}=1.4 M_{\rm sun}$. One predicts that $\lambda \simeq$ 24 km with the SLy4 model and $\lambda \simeq $ 15 km with the KIDS0-m*87 model. Both the KIDS0-m*87 and SLy4 models give $R_{\rm NS} \simeq 11.7$ km for the $1.4 M_{\rm sun}$ mass neutron stars, so they always have $\lambda \geq R_{\textrm{NS}}$ as shown in Fig.~\ref{fig8}. It means that both KIDS0-m*87 and SLy4 models support the small possibility of neutrino trapping within NSs. In the KIDS0, KIDS-A, and KIDS-B models, for stars heavier than the Sun, $\lambda$ is always shorter than $R_{\rm NS}$, so the neutrino emission can be delayed 
and thus have a non-negligible effect on the thermal evolution of NS.

%%%%%FIG9%%%%%
\begin{figure}[t]
  \begin{center}
    \includegraphics[width=0.60\textwidth]{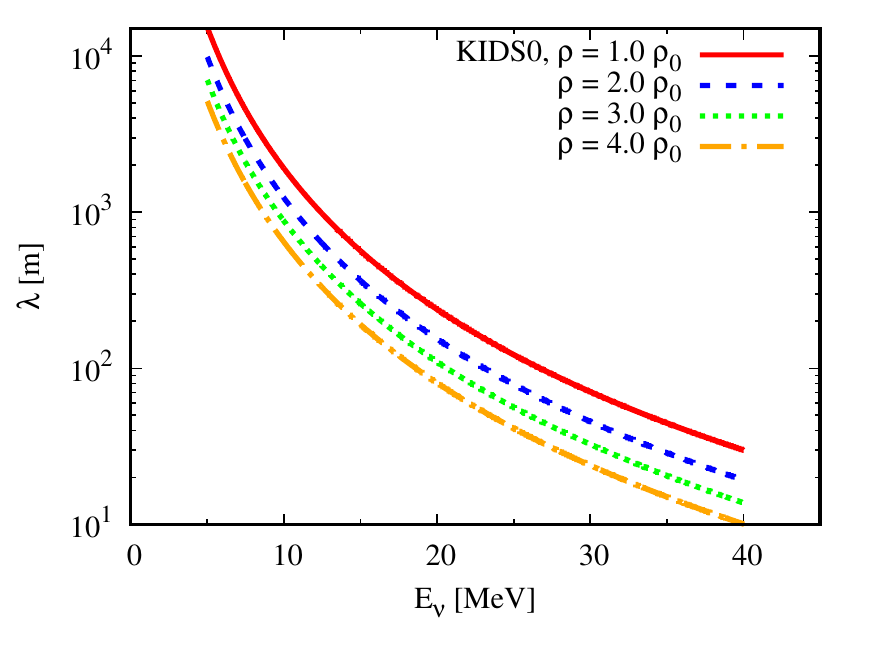}
  \end{center}
  \caption{Neutrino mean free path for the KIDS0 model with $M_n^* /M \gtrsim 1$ and soft symmetry energy as a function of the initial neutrino energy $E_{\nu}$.}
  \label{fig9}
\end{figure}

Finally, in Fig.~\ref{fig9} we show the NMFP for the KIDS0 model as a function of the initial neutrino energy $E_\nu$ for different densities. It is clearly seen that the NMFP decreases as the density and initial neutrino energy increase, which is consistent with the result obtained in the functionals of the Brussels-Montreal Skyrme (BSk17 and BSk18) models~\cite{Pastore:2014yua} at $\rho = 1.0~\rho_0$. However, in the present work, we also predict the NMFP for higher densities up to $\rho = 4.0~\rho_0$. Note that higher neutrino energy can be relevant not only for the NS but also for supernovae, i.e., $E_\nu =$ 30 MeV is a typical neutrino energy for the core-collapse supernova. Short NMFP can alter significantly the neutrino opacities by neutral-current scatterings and especially by the charged-current scattering in the hot and dense supernova matter \cite{prc2017}.
This strongly urges accurate determination of the nucleon effective masses in dense nuclear matter.

%--------------------------------------------
\section{Summary and conclusion} \label{sec:summary}
%--------------------------------------------
In the present work, we have investigated the nuclear symmetry energy, the proton, and neutron effective masses, the NS M-R relations, proton and neutron fractions, and NMFP using various KIDS-EDF models. The KIDS-EDF model has been widely and successfully applied to describe the properties of finite nuclei. In this work, we extend the applications of the KIDS-EDF model to the neutron star matter and the neutrino interaction with NS matter constituents using the LRA. We analyze the implications of the nucleon effective masses and symmetry energy for the NS properties through the M-R relations and neutrinos' interaction with the NS matter constituents as well as the implications of the NS M-R relations with the NMFP. 

We find that the nuclear symmetry energy for the KIDS0, KIDS0-m*87, and SLy4 models have quite similar behavior over the range of the nuclear densities, yielding soft nuclear symmetry energy. The KIDS-A and KIDS-B models have stiff nuclear symmetry energy. We also find that the KIDS-A model has the stiffest nuclear symmetry energy at $\rho > \rho_0$ amongst the considered models. Similar stiffness of the nuclear symmetry energy for the KIDS0, KIDS0-m*87, and SLy4 models leads to a similar result for the NS M-R relations, as shown in Fig.~\ref{fig3}(a). Results for the properties of NSs (M-R relations) for all the KIDS-EDF models are in excellent agreement with the recent observations of PSR J0348+0432~\cite{Antoniadis:2013pzd}, PSR J0740+6620~\cite{NANOGrav:2019jur}, and PSR J1614-2230~\cite{Demorest:2010bx}, predicting the maximum mass of a NS is larger than $2M_{\rm sun}$, as well as NICER observation results of radius $R_{\textrm{NS}} =$ 12.35 $\pm$ 0.75 km for a NS mass $M_{\textrm{NS}} =$ 2.08 $M_\mathrm{sun}$~\cite{Miller:2021qha}.

We find that the dominant contribution to the magnitude of the total DCRS of the neutrino is given by the DCRS of the neutrino with neutrons for KIDS0, KIDS-A, and KIDS-B models. 
The shape of the total DCRS of the neutrino for all the KIDS-EDF models depends on both shapes of the DCRS of the neutrino with protons and neutrons. The total cross sections for the KIDS0, KIDS-A, and KIDS-B models increase as the density increases, but for KIDS0-m*87 and SLy4 models the total cross sections of the neutrino are almost the same as the density increases. Amongst the KIDS-EDF models, the highest total cross section of the neutrino is given by the KIDS0 model. It is clearly understood taht the effective masses of neutrons for the KIDS0, KIDS-A, and KIDS-B models with $M_n^*/M \gtrsim 1$, being quite similar up to $\rho \simeq 2.0~\rho_0$, leading to similar prediction results for the DCRS of the neutrino.

More interestingly, we also find that the DCRSs of the neutrino for the KIDS0, KIDS-A, and KIDS-B models with $M_n^*/M\gtrsim 1$ are bigger than those obtained for the KIDS0-m*87 and SLy4 models with $M_n^*/M \lesssim 1$ at $\rho \simeq 1.0~\rho_0$. The ranges of $q_0$ for the KIDS0, KIDS-A, and KIDS-B models are smaller than those obtained for the KIDS0-m*87 and SLy4 models. At higher densities, $\rho \simeq 4.0~\rho_0$ differences in the DCRS of the neutrino among the KIDS-EDF models are much more pronounced because the effective mass differences in the models become significant.

In the results for the NMFP, we find the NMFPs for the KIDS-A and KIDS-B models are quite similar at all the baryon density regimes up to $6 \rho_0$. 
This behavior is followed by the KIDS0 model up to $\rho \simeq 3.0~\rho_0$ and the NMPF of the KIDS0 model decreases faster than 
those obtained for the KIDS-A and KIDS-B models. The NMFP predictions for the KIDS0-m*87 and SLy4 models are much higher compared with those 
obtained for the KIDS0, KIDS-A, and KIDS-B models.

The result shows that the NMFP is most strongly correlated to the neutron effective mass $M^*_n$. There are two dominant sources that control the behavior of $M^*_n$.
The most direct and evident effect comes from the isoscalar and isovector effective masses. It is shown that not only the values of the isoscalar and isovector effective masses but also their relative magnitudes are critical for the density dependence of $M^*_n$. 
Such a dependence on the isoscalar and isovector effective masses appears in the same way 
in the neutrino scattering with finite nuclei \cite{nuA2022}. Therefore the correct description of the neutrino interaction in/with nuclear many-body systems demands the exact determination of both isoscalar and isovector effective masses from either theory or experiment. Another quantity critical to $M^*_n$ is the symmetry energy. Its effect is not as direct as the isoscalar and isovector effective masses but is embedded in the neutron to proton asymmetry $\delta$. Symmetry energy is the main source of the different particle fractions between the models in Fig.~3. Since isoscalar and isovector effective masses are similar in the KIDS0, KIDS-A, and KIDS-B models, the behavior of $M^*_n$ in the KIDS0 model is different from those of the KIDS-A and KIDS-B models at $\rho \gtrsim 3 \rho_0$ and is attributed to the difference in the particle fraction. Accurate measurement of the symmetry energy from sub to supra saturation densities is another essential requirement to refine our understanding of the role of neutrinos in supernova explosions and neutron star cooling.

Summarizing the result for the $\lambda$-$R_{\textrm{NS}}$ relations, the KIDS0, KIDS-A, and KIDS-B models show that $\lambda \lesssim R_{\textrm{NS}}$, indicating that they could lead to neutrino trapping in NS and thus the delayed NS cooling. The KIDS0-m*87 and SLy4 models always have $\lambda \gtrsim R_{\textrm{NS}}$, as clearly shown in Fig.~\ref{fig8}, thus they support a small possibility of neutrino trapping within the NS.

We finally find that the NMFP decreases as the density and initial neutrino energy increase, which is consistent with those obtained in the functionals of the Brussels-Montreal Skyrme (BSk17 and BSk18) models at $\rho = \rho_0$~\cite{Pastore:2014yua}.

\section*{Acknowledgments}
This work was supported by the National Research Foundation of Korea (NRF) Grant No.~2018R1A5A1025563. The works of S.i.N. and P.T.P.H. were partially supported by the NRF Grant No.~2022R1A2C1003964. The work of C.H.H. was also partially supported by the NRF Grant No.~2020R1F1A1052495.

\end{document}